\documentclass{article}

\usepackage{etex}

\usepackage{authblk}

\usepackage{comment}
\usepackage{tikz}
\usetikzlibrary{matrix,arrows,decorations.pathmorphing,patterns,positioning}
\usepackage{amssymb}
\usepackage{ifthen}
\usepackage{color}
\usepackage{colortbl}
\usepackage{rotating}
\usepackage{multirow}
\usepackage{multicol}
\usepackage{url}
\usepackage{arydshln}

\usepackage{pgfplots}

\usetikzlibrary{external}
\usepackage{wrapfig}
\usepackage{subfigure}
\usepackage{array}
\usepackage{framed}

\usepackage{balance}

\usepackage{epsfig}
\usepackage{amsmath}
\usepackage{amsfonts}
\usepackage{amssymb}

\usepackage{algorithmic}

\usepackage{algorithm}

\usepackage[normalem]{ulem}

\usepackage{ifthen}
\usepackage{boxedminipage}
\usepackage{fancyvrb}

\usepackage{soul}
\usepackage{graphicx}
\usepackage{graphics}
\usepackage{moreverb}
\usepackage{rotating}
\usepackage{enumerate}

\usepackage{tabularx,booktabs}

\usetikzlibrary{arrows}

\usepackage{xspace}

\usepackage{stmaryrd}

\usepackage{bm}

\usepackage[absolute]{textpos}
\TPMargin{5pt}

\usepackage{listings}
\usepackage{xcolor}
\usepackage{fancyvrb}
\fvset{tabsize=2}
\definecolor{mygreen}{rgb}{0,0.6,0}
\pgfplotscreateplotcyclelist{mycolorlist}{%
blue,every mark/.append style={fill=blue!80!black},mark=*\\%
red,every mark/.append style={fill=red!80!black},mark=square*\\%
brown!60!black,every mark/.append style={fill=brown!80!black},mark=otimes*\\%
black,mark=star\\%
blue,every mark/.append style={fill=blue!80!black},mark=diamond*\\%
red,densely dashed,every mark/.append style={solid,fill=red!80!black},mark=*\\%
brown!60!black,densely dashed,every mark/.append style={
solid,fill=brown!80!black},mark=square*\\%
black,densely dashed,every mark/.append style={solid,fill=gray},mark=otimes*\\%
blue,densely dashed,mark=star,every mark/.append style=solid\\%
red,densely dashed,every mark/.append style={solid,fill=red!80!black},mark=diamond*\\%
}

\definecolor{colorbrewer1}{RGB}{228,26,28}
\definecolor{colorbrewer2}{RGB}{55,126,184}
\definecolor{colorbrewer3}{RGB}{77,175,74}
\definecolor{colorbrewer4}{RGB}{152,78,163}
\definecolor{colorbrewer5}{RGB}{255,127,0}
\definecolor{colorbrewer7}{RGB}{166,86,40}
\definecolor{colorbrewer8}{RGB}{247,129,191}
\definecolor{colorbrewer9}{RGB}{153,153,153}

\definecolor{darkgreen}{rgb}{0,0.5,0}

\definecolor{darkred}{rgb}{0.44,0,0}
\definecolor{darkgreen}{rgb}{0,0.44,0}
\definecolor{darkblue}{rgb}{0,0,0.44}
\definecolor{enrique}{rgb}{0,0,0}

\pgfplotscreateplotcyclelist{jianyu2color}{
	colorbrewer1, mark=*,\\
	colorbrewer2, mark=square*,\\
	colorbrewer3, mark=triangle*,\\
	colorbrewer4, mark=x,\\
	colorbrewer5, mark=*,\\
	colorbrewer7, mark=triangle*,\\
	colorbrewer8, mark=+,\\
}

\pgfplotscreateplotcyclelist{jianyu0color}{%
ultra thick,colorbrewer1,every mark/.append style={fill=colorbrewer1},mark=triangle*\\%
orange,every mark/.append style={fill=orange},mark=square*\\%
cyan,every mark/.append style={fill=cyan},mark=otimes*\\%
red!70!white,mark=star\\%
lime!80!black,every mark/.append style={fill=lime},mark=diamond*\\%
red,densely dashed,every mark/.append style={solid,fill=red!80!black},mark=*\\%
thick,yellow!60!black,every mark/.append style={solid,fill=yellow!80!black},mark=square*\\%
purple,every mark/.append style={solid,fill=gray},mark=otimes*\\%
blue,densely dashed,mark=star,every mark/.append style=solid\\%
red,densely dashed,every mark/.append style={solid,fill=red!80!black},mark=diamond*\\%
thick,colorbrewer4,mark=x\\%
blue,every mark/.append style={fill=blue!80!black},mark=*\\%
red,every mark/.append style={fill=red!80!black},mark=square*\\%
brown!60!black,every mark/.append style={fill=brown!80!black},mark=diamond*\\%
thick,darkgreen,mark=star\\%
blue,every mark/.append style={fill=blue!80!black},mark=diamond*\\%
red,densely dashed,every mark/.append style={solid,fill=red!80!black},mark=*\\%
brown!60!black,densely dashed,every mark/.append style={
solid,fill=brown!80!black},mark=square*\\%
darkred,densely dashed,every mark/.append style={solid,fill=gray},mark=otimes*\\%
blue,densely dashed,mark=oplus,every mark/.append style=solid\\%
colorbrewer5,densely dashed,every mark/.append style={solid,fill=colorbrewer5},mark=diamond*\\%
darkgreen,densely dashed,mark=+,every mark/.append style=solid\\%
thick,colorbrewer7,mark=triangle*\\%
thick,black\\%
}

\pgfplotscreateplotcyclelist{jianyucolor}{%
ultra thick,colorbrewer1,every mark/.append style={fill=colorbrewer1},mark=triangle*\\%
orange,every mark/.append style={fill=orange},mark=square*\\%
cyan,every mark/.append style={fill=cyan},mark=otimes*\\%
red!70!white,mark=star\\%
lime!80!black,every mark/.append style={fill=lime},mark=diamond*\\%
red,densely dashed,every mark/.append style={solid,fill=red!80!black},mark=*\\%
thick,yellow!60!black,every mark/.append style={solid,fill=yellow!80!black},mark=square*\\%
purple,every mark/.append style={solid,fill=gray},mark=otimes*\\%
blue,densely dashed,mark=star,every mark/.append style=solid\\%
red,densely dashed,every mark/.append style={solid,fill=red!80!black},mark=diamond*\\%
thick,colorbrewer4,mark=x\\%
blue,every mark/.append style={fill=blue!80!black},mark=*\\%
red,every mark/.append style={fill=red!80!black},mark=square*\\%
brown!60!black,every mark/.append style={fill=brown!80!black},mark=diamond*\\%
thick,darkgreen,mark=star\\%
blue,every mark/.append style={fill=blue!80!black},mark=diamond*\\%
red,densely dashed,every mark/.append style={solid,fill=red!80!black},mark=*\\%
brown!60!black,densely dashed,every mark/.append style={
solid,fill=brown!80!black},mark=square*\\%
darkred,densely dashed,every mark/.append style={solid,fill=gray},mark=otimes*\\%
blue,densely dashed,mark=oplus,every mark/.append style=solid\\%
colorbrewer5,densely dashed,every mark/.append style={solid,fill=colorbrewer5},mark=diamond*\\%
darkgreen,densely dashed,mark=+,every mark/.append style=solid\\%
thick,colorbrewer7,mark=triangle*\\%
thick,black,densely dashed\\%
thick,black\\%
}

\pgfplotscreateplotcyclelist{jianyu3color}{%
black,densely dashed\\%
black\\%
colorbrewer1,every mark/.append style={fill=colorbrewer1},mark=triangle*\\%
orange,every mark/.append style={fill=orange},mark=square*\\%
cyan,every mark/.append style={fill=cyan},mark=otimes*\\%
red!70!white,mark=star\\%
brown!60!black,every mark/.append style={fill=brown!80!black},mark=otimes*\\%
purple,every mark/.append style={solid,fill=gray},mark=otimes*\\%
very thick, blue,every mark/.append style={fill=lime},mark=diamond*\\%
very thick, darkgreen,mark=+,every mark/.append style=solid\\%
}

\pgfplotscreateplotcyclelist{jianyu4color}{%
black,densely dashed,mark=triangle*\\%
black,mark=triangle*\\%
colorbrewer1,densely dashed,mark=triangle*\\%
thick,colorbrewer1,mark=triangle*\\%
black,densely dashed,mark=square*\\%
black,mark=square*\\%
blue,densely dashed,mark=square*\\%
thick,blue,mark=square*\\%
black,densely dashed,mark=*\\%
black,mark=*\\%
darkgreen,densely dashed,mark=*\\%
thick,darkgreen,mark=*\\%
orange,mark=square\\%
cyan,mark=otimes\\%
purple,mark=star\\%
}

\pgfplotscreateplotcyclelist{jianyu5color}{%
black,densely dashed\\%
black\\%
thick,densely dashed,colorbrewer1,every mark/.append style={fill=colorbrewer1},mark=triangle*\\%
orange,every mark/.append style={fill=orange},mark=square*\\%
cyan,every mark/.append style={fill=cyan},mark=otimes*\\%
red!70!white,mark=star\\%
brown!60!black,every mark/.append style={fill=brown!80!black},mark=otimes*\\%
purple,every mark/.append style={solid,fill=gray},mark=otimes*\\%
blue,every mark/.append style={fill=lime},mark=diamond*\\%
darkgreen,mark=+,every mark/.append style=solid\\%
}

\newcommand{\GFLOPS}{GFLOPS}

\newcommand{\ABCFMM}{\textbf{ABC \mbox{FMM}}}
\newcommand{\ABXFMM}{\textbf{AB \mbox{FMM}}}
\newcommand{\XXXFMM}{\textbf{Naive \mbox{FMM}}}

\newcommand{\figref}[1]{Figure~\ref{#1}}

\newcommand{\strassen}{\mbox{\sc Strassen}}

\newcommand{\GOTO}{{\sc GotoBLAS}}
\newcommand{\BLIS}{{\sc BLIS}}

\newcommand{\rr}[0]{\textsuperscript{\textregistered}\xspace}

\newcommand{\gemm}{{\sc gemm}\xspace}
\newcommand{\ngemm}{{\sc gemm}\xspace}
\newcommand{\dgemm}{{\sc dgemm}}

\newcommand{\FMM}{{\sc FMM}}

\newcommand{\fromto}[2]{{\color{black} #2}}

\newcolumntype{I}{!{\vrule width 1.5pt}}
\newlength\savedwidth
\newcommand\whline{\noalign{\global\savedwidth\arrayrulewidth
                            \global\arrayrulewidth 1.5pt}%
           \hline
           \noalign{\global\arrayrulewidth\savedwidth}}

\newcommand{\DivisorML}{ \widetilde{M_L} }
\newcommand{\DivisorNL}{ \widetilde{N_L} }
\newcommand{\DivisorKL}{ \widetilde{K_L} }
\newcommand{\OTIMESU}{ \bigotimes \! U }
\newcommand{\OTIMESV}{ \bigotimes \! V }
\newcommand{\OTIMESW}{ \bigotimes \! W }

\newcommand{\PRODRL}{ R_L }

\newcommand{\SQUARE}{$ m\!\!=\!\! k \!\!=\!\! n $}
\newcommand{\RANKK}{$ m \!\!=\!\! n \!\!=\!\! 14400 $, $ k $ varies}
\newcommand{\FIXK}{$ k \!\!=\!\! 1024 $, $ m \!\! =\!\! n $ vary}

\newcommand{\mycircle}[1]{{\textcircled{\raisebox{-0.9pt}{#1}}}}

\newcommand*{\affaddr}[1]{#1} 
\newcommand*{\affmark}[1][*]{\textsuperscript{#1}}
\newcommand*{\email}[1]{\texttt{#1}}

\newcommand{\FromTo}[2]{{\color{black} #2}}

\date{
November 3, 2016
}

\begin{document}

\title{
\Large Generating Families of Practical Fast Matrix Multiplication Algorithms
\\[0.2in]
\large FLAME Working Note \#82
}


\author{
Jianyu Huang\affmark[*]\affmark[\dag],
Leslie Rice\affmark[*]\affmark[\dag],
Devin A. Matthews\affmark[\dag],
Robert A. van de Geijn\affmark[*]\affmark[\dag]\\
\affaddr{\affmark[*]Department of Computer Science}\\
\affaddr{\affmark[\dag]Institute for Computational Engineering and Sciences}\\
\affaddr{The University of Texas at Austin, Austin, TX 78712}\\
\email{{\tt \{jianyu@cs.}, {\tt leslierice}, {\tt dmatthews}, {\tt rvdg@cs.\}utexas.edu}}
}


\newcommand{\NoShow}[1]{}


\maketitle

\begin{abstract}

Matrix multiplication (GEMM) is a core operation to numerous scientific applications. Traditional implementations of Strassen-like fast matrix multiplication (FMM) algorithms often do not perform well except for very large matrix sizes, due to the increased cost of memory movement, which is particularly noticeable for non-square matrices. Such implementations also require considerable workspace and modifications to the standard BLAS interface. We propose a code generator framework to automatically implement a
large family of FMM algorithms suitable for multiplications of arbitrary matrix sizes and shapes.
By representing FMM with a triple of matrices $ \llbracket U, V, W \rrbracket $ that capture the linear combinations of submatrices that are formed, we can use the Kronecker product to define a multi-level representation of Strassen-like algorithms. Incorporating the matrix additions that must be performed for Strassen-like algorithms into the inherent packing and micro-kernel operations inside GEMM avoids extra workspace and reduces the cost of memory movement. Adopting the same loop structures as high-performance GEMM implementations allows parallelization of all FMM algorithms with simple but efficient data parallelism without the overhead of task parallelism. We present a simple performance model for general FMM algorithms and compare
actual performance of 20+ FMM algorithms to modeled predictions. Our implementations demonstrate a performance benefit over conventional GEMM on
\FromTo{single-core, multi-core, and many-core (second-generation Intel Xeon Phi) systems.}{single core and multi-core systems.}
This study shows that Strassen-like fast matrix multiplication can be incorporated into libraries for practical use.


\end{abstract}

\section{Introduction}
\label{sec:introduction} \label{s:intro}

\FromTo{}{Three recent advances have revived interest in the practical implementation of Strassen's algorithm (\strassen) and similar Fast Matrix Multiplication  (\FMM{}) algorithms.  The first~\cite{Benson15} is a systematic way in which new FMM algorithms can be identified, building upon conventional calls to the BLAS matrix-matrix multiplication \ngemm\ routine.
That work incorporated a code generator, due to the number of algorithms that are identified and the complexity of exploiting subexpressions encountered in the linear combinations of submatrices.
Parallelism was achieved through a combination of task parallelism and parallelism within the BLAS.  The second~\cite{Strassen:SC16} was the insight that the BLAS-like Library Instantiation Software (BLIS) framework exposes basic building blocks that allow the linear combinations of submatrices in \strassen\ to be incorporated into the packing and/or computational micro-kernels already existing in the BLIS \ngemm\ implementation.
Parallelism in that work mirrored the highly effective data parallelism that is part of BLIS.
Finally, the present work also extends insights on how to express multiple levels of \strassen\ in terms of Kronecker products~\cite{KroneckerStrassen} to multi-level FMM algorithms, facilitating a code generator for all methods from~\cite{Benson15} (including \strassen), in terms of the building blocks created for~\cite{Strassen:SC16}, but allowing different FMM algorithms to be used for each level.  Importantly and unique to this work, the code generator also yields performance models that are accurate enough to guide the choice of a FMM implementation as a function of problem size and shape, facilitating the creation of poly-algorithms~\cite{TonyPoly}.
\FromTo{Performance results from single core, multi-core, and many-core architectures support the theoretical insights.}{Performance results from single core and multi-core shared memory system support the theoretical insights.}

We focus on the special case of \ngemm\ given by
$ C := C + A B $.  Extending the ideas to the more general case of \ngemm\ is straightforward.
}

\section{Background} \label{s:background}

We briefly summarize how the BLIS~\cite{BLIS1} framework implements \ngemm\ before reviewing recent results~\cite{Strassen:SC16} 
on how \strassen\ can exploit insights that underlie this framework.

\subsection{High-performance implementation of standard \gemm}

\NoShow{
(General) matrix-matrix multiplication (\ngemm) is supported in the level-3 BLAS~\cite{BLAS3} interface as
\begin{quote}
	\footnotesize
	\tt DGEMM( \begin{tabular}[t]{@{}l}
			transa, transb, m, n, k, alpha, \\
	\tt ~~~~~~~A, lda, B, ldb, beta, C, ldc )
	\end{tabular}
\end{quote}
where we focus on double precision arithmetic and data. 
}

\begin{figure*}[tb!]
~
\vspace{-0.60in}
\begin{center}
\includegraphics[width=1.06\textwidth]{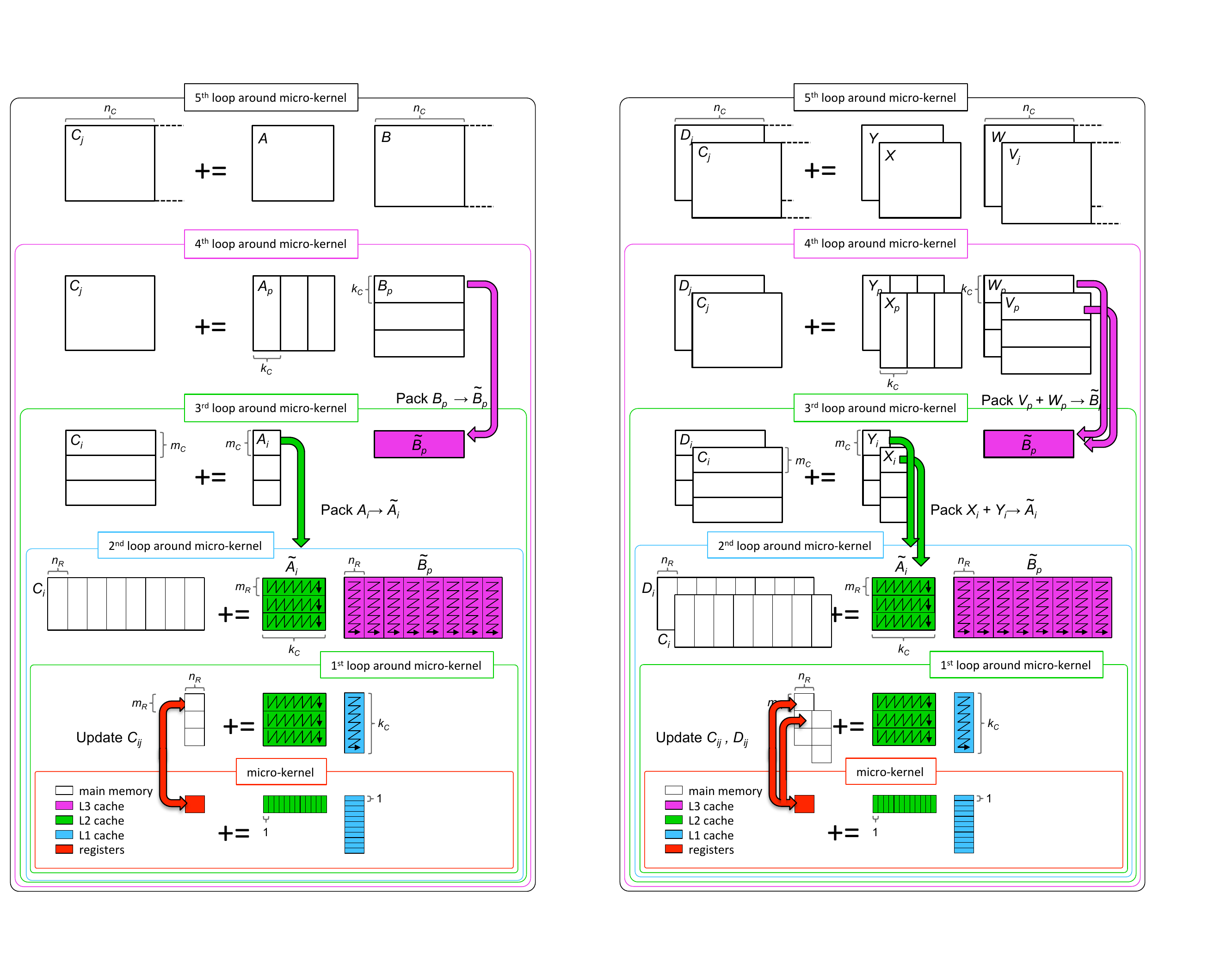}
\end{center}
\vspace{-0.55in}
\caption{
	\FromTo{}{Figure from~\cite{Strassen:SC16} (used with permission from authors).}
	Left: Illustration of the BLIS implementation of the \GOTO{} \gemm\  algorithm.  All computation is cast in terms of a micro-kernel that is highly optimized.    Right: modification that implements the representative computation $ M = ( X + Y)( V+W); C +\!\!= M; D +\!\!= M $ of each row of computations in (\ref{eqn:allops}).
\FromTo{}{$ X $, $ Y $ are submatrices of $ A $; $ V $, $ W $ are submatrices of $ B $; $ C $, $ D $ are submatrices of the original matrix $ C $; $ M $ is the intermediate matrix product. Note that the packing buffers $ \widetilde A_i $ and $ \widetilde B_p $ stay in cache. }
}
\label{fig:side_by_side}
\end{figure*}

Key to high performance implementations of \gemm\ is
the partitioning of operands in order 
to (near-)optimally reuse data in the various levels of memory.
\figref{fig:side_by_side}(left) illustrates how BLIS implements the \GOTO{} \cite{Goto:2008:AHP} approach.
Block sizes $\{m_C, n_C, k_C \} $ are chosen so that submatrices fit in the various caches while $ \{ m_R, n_R \} $ relate to submatrices in registers that contribute to $ C $.  These parameters can be analytically determined~\cite{BLIS4}.
To improve data locality, row panels $ B_p $ that fit in the L3 cache  
are ``packed'' into contiguous memory, yielding $ \widetilde B_p $.  
For similar reasons, 
blocks $ A_i $ that fit in the L2 cache are packed into buffer $ \widetilde A_i $.  

\subsection{High-performance implementations of \strassen}

If one partitions the three operands into quadrants,
\begin{equation}
\label{eqn:strassenpart}
X =
\left( \begin{array}{c | c}
X_{0} & X_{1} \\ \hline
X_{2} & X_{3}
\end{array}\right)
\quad \mbox{for $ X \in \{ A, B, C\} $}
\end{equation}
then it can be shown that the operations
\begin{equation}
{
\begin{array}{l @{\hspace{1pt}} c @{\hspace{1pt}} l l r }
M_0 &=&  ( A_{0} + A_{3} ) ( B_{0} + B_{3} );
&
C_{0} += M_0;  & C_{3} += M_0;  \\
M_1 &=&  ( A_{2} + A_{3} ) B_{0};
&
C_{2} += M_1 ;  & C_{3} -= M_1 ; \\
M_2 &=&  A_{0} ( B_{1} - B_{3} );
&
C_{1} += M_2 ;  & C_{3} += M_2 ;
\\
M_3 &=&  A_{3}( B_{2} - B_{0} );
&
C_{0} += M_3 ;  & C_{2} += M_3 ;
\\
M_4 &=&  ( A_{0} + A_{1}) B_{3};
&
C_{1} +=  M_4 ;  &  C_{0} -= M_4;
\\
M_5&=&  (A_{2} - A_{0} )( B_{0} + B_{1} );
&
C_{3} += M_5;
\\
M_6&=&  (A_{1} - A_{3} )( B_{2} + B_{3} );
&
C_{0} += M_6 ;
\end{array}
}
\label{eqn:allops}
\end{equation} %
compute $ C := A B + C $, but with seven (sub)matrix multiplications, reducing the cost by a factor of $7/8$ (ignoring a lower order number of extra additions).  If all matrices are square and of size $ n \times n $, classical \strassen\ exploits this recursively, reducing the cost for \gemm\ to $ O( n^{2.807} ) $.

Only a few levels of the recursion are exploited in practice because the cost of extra additions and extra memory movements quickly offsets the reduction in floating point operations.  Also, \strassen\ is known to become more numerically unstable particularly when more than two levels of recursion are employed~\cite{HighamBook,DemmelStrassenStable2007,Ballard15}.

In~\cite{Strassen:SC16}, captured in \figref{fig:side_by_side}(right), it was noted that the additions of the
submatrices of $ A $ and $ B $ can be incorporated into the packing into buffers $ \widetilde A_i $ and $ \widetilde B_p $, avoiding extra memory movements.  In addition, once a submatrix that contributes to $ C $ is computed in the registers, it can be directly added to the appropriate parts of multiple submatrices of $ C $, thus avoiding the need for temporary matrices $ M_r $, again avoiding extra memory movements. As demonstrated in~\cite{Strassen:SC16}, this makes the method practical for smaller matrices and matrices of special shape (especially rank-k updates, where $ k $ is relatively small).

\section{Fast Matrix Multiplication Algorithms} \label{s:method}

We now present the basic idea that underlies families of FMM algorithms
and how to \FromTo{use}{generalize} one-level formula for multi-level FMM utilizing Kronecker products and recursive block storage indexing.

\renewcommand{\bm}[1]{#1}

\subsection{One-level fast matrix multiplication algorithms}

In~\cite{Benson15}, the theory of tensor contractions is used to find a large number of FMM algorithms. 
In this subsection, we use the output (the resulting algorithms) of their approach.

\NoShow{FMM algorithms are defined by how one level of recursion recasts the required computation in terms of matrix multiplications with linear combinations of submatrices of the operands $ A $ and $ B $, after which a linear combination of the results contributes to submatrices of $ C $.}

Generalizing the partitioning for \strassen,
consider $ C := C + AB $, where $ C $, $ A $, and $ B $ are $ m \times n $, $ m \times k $, and $ k \times n $ matrices, respectively.
\cite{Benson15} defines a $ \langle \widetilde m, \widetilde k, \widetilde n \rangle $ algorithm by partitioning
{
\setlength{\arraycolsep}{1pt}
\[
\begin{array}{c}
C = \left( \begin{array}{c | c | c }
C_{0} & \cdot  \cdot  \cdot & C_{\widetilde n-1} \\ \hline
\vdots & & \vdots \\ \hline
C_{(\widetilde m-1) \widetilde n} & \cdot  \cdot  \cdot & C_{\widetilde m \widetilde n-1}
\end{array} \right),
A = \left( \begin{array}{c | c | c }
A_{0} & \cdot  \cdot  \cdot & A_{\widetilde k - 1} \\ \hline
\vdots & & \vdots \\ \hline
A_{(\widetilde m-1)\widetilde k} & \cdot  \cdot  \cdot & A_{\widetilde m \widetilde k - 1}
\end{array} \right), 
\mbox{and }
B = \left( \begin{array}{c | c | c }
B_{0} & \cdot  \cdot  \cdot & B_{\widetilde n-1} \\ \hline
\vdots & & \vdots \\ \hline
B_{(\widetilde k-1) \widetilde n} & \cdot  \cdot  \cdot & B_{\widetilde k \widetilde n - 1 }
\end{array} \right)
\end{array}
\]}
Note that $ A_i $, $ B_j $, and $ C_p $ are the submatrices of $ A $, $ B $ and $ C $, with a single index in the row major order.
Then, $ C := C + A B $ is computed by,

\text{for } $ r = 0,..., R-1 $,
\begin{equation}
\begin{array}{l}
M_r :=
\left( \sum\limits_{i=0}^{\widetilde m \widetilde k-1} { u_{ir}  A_i } \right)
\times
\left( \sum\limits_{j=0}^{\widetilde k \widetilde n-1} { v_{jr}  B_j } \right); \\
C_p +=  w_{pr} M_r~( p = 0,..., \widetilde{m}\widetilde{n}-1)
\end{array}
\label{e:fmm_1}
\end{equation}
where ($ \times $) is a matrix multiplication that can be done recursively,
$ u_{ir} $, $ v_{jr} $, and $ w_{pr} $ are entries of a $ (\widetilde{m}\widetilde{k}) \times R $ matrix $ {U} $,
a $ (\widetilde{k}\widetilde{n}) \times R $ matrix $ V $, and
a $ (\widetilde{m}\widetilde{n}) \times R $ matrix $ W $,
respectively.
Therefore, the classical matrix multiplication which needs $ \widetilde m \widetilde k \widetilde n $ submatrix multiplications can be
completed with $ R $ submatrix multiplications.
The set of coefficients that determine the $ \langle \widetilde{m}, \widetilde{k}, \widetilde{n} \rangle $ algorithm is denoted as
$ \llbracket U, V, W \rrbracket $.

For example, assuming that $ m $, $ n $, and $ k$ are all even, one-level \strassen{} has
$ \langle 2, 2, 2 \rangle $ partition dimensions and, given the partitioning in (\ref{eqn:strassenpart}) and computations in~(\ref{eqn:allops}),{\setlength{\arraycolsep}{2pt}
\begin{equation}
{
	\setlength{\arraycolsep}{1.5pt}
\llbracket 
\left( \begin{array}{@{}r r r r r r r r r@{}}
\phantom{-}1 & \phantom{-}0 & \phantom{-}1 & \phantom{-}0 & \phantom{-}1 & -1 & \phantom{-}0 \\
0 & 0 & 0 & 0 & 1 & 0  & 1 \\
0 & 1 & 0 & 0 & 0 & 1  & 0 \\
1 & 1 & 0 & 1 & 0 & 0  & -1 \\
\end{array}\right),
\left( \begin{array}{@{}r r r r r r r r r@{}}
\phantom{-}1 & \phantom{-}1 & \phantom{-}0  & -1 & \phantom{-}0 & \phantom{-}1 & \phantom{-}0 \\
0 & 0 & 1  & 0  & 0 & 1 & 0 \\
0 & 0 & 0  & 1  & 0 & 0 & 1 \\
1 & 0 & -1 & 0  & 1 & 0 & 1 \\
\end{array}
\right),
\left( \begin{array}{@{}r r r r r r r r r@{}}
\phantom{-}1 & \phantom{-}0 & \phantom{-}0 & \phantom{-}1 & -1 & \phantom{-}0 & \phantom{-}1 \\
0 & 0 & 1 & 0 & 1 & 0 & 0 \\
0 & 1 & 0 & 1 & 0 & 0 & 0 \\
1 & -1 & 1 & 0 & 0 & 1 & 0 \\
\end{array}
\right) 
\rrbracket
}
\label{eqn:stra_coeffs}
\end{equation}
specifies 
 $ \llbracket {U}, {V}, {W} \rrbracket $ for
one-level \strassen{}.}

\begin{figure}[!t]
\centering
{
\begin{tabular}{ c | c c c | r | r | r |r | r}
\whline
 \multirow{3}{*}{$ \langle \widetilde m, \widetilde k, \widetilde n \rangle $}  & \multirow{3}{*}{Ref.} & \multirow{3}{*}{$\widetilde m  \widetilde k  \widetilde n $}  & \multirow{3}{*}{$ R $} & \multicolumn{5}{c} {Speedup ($\%$)}\\
\cline{5-9}
 & & & & \multirow{2}{*}{Theory} & \multicolumn{2}{c|}{Practical \#1} & \multicolumn{2}{c}{Practical \#2} \\
\cline{6-9}
 & & & &                              & Ours & \multicolumn{1}{c|}{\cite{Benson15}}  &  Ours & \multicolumn{1}{c}{\cite{Benson15}} \\
\hline
$ \langle 2, 2, 2 \rangle $ & \cite{Strassen}    & $  8 $ & $ 7 $ & $ 14.3  $ & $ 11.9        $~&~$ \text{-}3.0     $ & $ 13.1 $~&~$ 13.1 $\\
$ \langle 2, 3, 2 \rangle $ & \cite{Benson15}    & $ 12 $ & $ 11 $ & $ 9.1  $ & $ 5.5         $~&~$ \text{-}13.1    $ & $ 7.7  $~&~$ 7.7  $\\
$ \langle 2, 3, 4 \rangle $ & \cite{Benson15}    & $ 24 $ & $ 20 $ & $ 20.0 $ & $ 11.9        $~&~$ \text{-}8.0     $ & $ 16.3 $~&~$ 17.0 $\\
$ \langle 2, 4, 3 \rangle $ & \cite{Ballard15}   & $ 24 $ & $ 20 $ & $ 20.0 $ & $ 4.8         $~&~$ \text{-}15.3    $ & $ 14.9 $~&~$ 16.6 $\\
$ \langle 2, 5, 2 \rangle $ & \cite{Ballard15}   & $ 20 $ & $ 18 $ & $ 11.1 $ & $ 1.5         $~&~$ \text{-}23.1    $ & $ 8.6  $~&~$ 8.3  $\\
$ \langle 3, 2, 2 \rangle $ & \cite{Ballard15}   & $ 12 $ & $ 11 $ & $ 9.1  $ & $ 7.1         $~&~$ \text{-}6.6     $ & $ 7.2  $~&~$ 7.5  $\\
$ \langle 3, 2, 3 \rangle $ & \cite{Ballard15}   & $ 18 $ & $ 15 $ & $ 20.0 $ & $ 14.1        $~&~$ \text{-}0.7     $ & $ 17.2 $~&~$ 16.8 $\\
$ \langle 3, 2, 4 \rangle $ & \cite{Ballard15}   & $ 24 $ & $ 20 $ & $ 20.0 $ & $ 11.9        $~&~$ \text{-}1.8     $ & $ 16.1 $~&~$ 17.0 $\\
$ \langle 3, 3, 2 \rangle $ & \cite{Ballard15}   & $ 18 $ & $ 15 $ & $ 20.0 $ & $ 11.4        $~&~$ \text{-}8.1     $ & $ 17.3 $~&~$ 16.5 $\\
$ \langle 3, 3, 3 \rangle $ & \cite{smi2013}     & $ 27 $ & $ 23 $ & $ 17.4 $ & $ 8.6         $~&~$ \text{-}9.3     $ & $ 14.4 $~&~$ 14.7 $\\
$ \langle 3, 3, 6 \rangle $ & \cite{smi2013}     & $ 54 $ & $ 40 $ & $ 35.0 $ & $ \text{-}34.0$~&~$ \text{-}41.6    $ & $ 24.2 $~&~$ 20.1 $\\
$ \langle 3, 4, 2 \rangle $ & \cite{Benson15}    & $ 24 $ & $ 20 $ & $ 20.0 $ & $ 4.9         $~&~$ \text{-}15.7    $ & $ 16.0 $~&~$ 16.8 $\\
$ \langle 3, 4, 3 \rangle $ & \cite{smi2013}     & $ 36 $ & $ 29 $ & $ 24.1 $ & $ 8.4         $~&~$ \text{-}12.6    $ & $ 18.1 $~&~$ 20.1 $\\
$ \langle 3, 5, 3 \rangle $ & \cite{smi2013}     & $ 45 $ & $ 36 $ & $ 25.0 $ & $ 5.2         $~&~$ \text{-}20.6    $ & $ 19.1 $~&~$ 18.9 $\\
$ \langle 3, 6, 3 \rangle $ & \cite{smi2013}     & $ 54 $ & $ 40 $ & $ 35.0 $ & $ \text{-}21.6$~&~$ \text{-}64.5    $ & $ 19.5 $~&~$ 17.8 $\\
$ \langle 4, 2, 2 \rangle $ & \cite{Ballard15}   & $ 16 $ & $ 14 $ & $ 14.3 $ & $ 9.4         $~&~$ \text{-}4.7     $ & $ 11.9 $~&~$ 12.2 $\\
$ \langle 4, 2, 3 \rangle $ & \cite{Benson15}    & $ 24 $ & $ 20 $ & $ 20.0 $ & $ 12.1        $~&~$ \text{-}2.3     $ & $ 15.9 $~&~$ 17.3 $\\
$ \langle 4, 2, 4 \rangle $ & \cite{Ballard15}   & $ 32 $ & $ 26 $ & $ 23.1 $ & $ 10.4        $~&~$ \text{-}2.7     $ & $ 18.4 $~&~$ 19.1 $\\
$ \langle 4, 3, 2 \rangle $ & \cite{Ballard15}   & $ 24 $ & $ 20 $ & $ 20.0 $ & $ 11.3        $~&~$ \text{-}7.8     $ & $ 16.8 $~&~$ 15.7 $\\
$ \langle 4, 3, 3 \rangle $ & \cite{Ballard15}   & $ 36 $ & $ 29 $ & $ 24.1 $ & $ 8.1         $~&~$ \text{-}8.4     $ & $ 19.8 $~&~$ 20.0 $\\
$ \langle 4, 4, 2 \rangle $ & \cite{Ballard15}   & $ 32 $ & $ 26 $ & $ 23.1 $ & $ \text{-}4.2 $~&~$ \text{-}18.4    $ & $ 17.1 $~&~$ 18.5 $\\
$ \langle 5, 2, 2 \rangle $ & \cite{Ballard15}   & $ 20 $ & $ 18 $ & $ 11.1 $ & $ 7.0         $~&~$ \text{-}6.7     $ & $ 8.2  $~&~$ 8.5  $\\
$ \langle 6, 3, 3 \rangle $ & \cite{smi2013}     & $ 54 $ & $ 40 $ & $ 35.0 $ & $ \text{-}33.4$~&~$ \text{-}42.2    $ & $ 24.0 $~&~$ 20.2 $\\


\whline
\end{tabular}
}
\caption{Theoretical and practical speedup for various FMM algorithms. $ \widetilde m \widetilde k \widetilde n $ is the number of multiplication for classical matrix multiplication algorithm. $ R $ is the number of multiplication for fast matrix multiplication algorithm. Theoretical speedup is the speedup per recursive step.
Practical \#1 speedup is the speedup for one-level FMM comparing with \gemm{} when $ m = n = 14400, k = 480 $ (rank-k updates). Practical \#2 speedup is the speedup for one-level FMM comparing with \gemm{} when $ m = n = 14400, k = 12000 $ (approximately square). We report the practical speedup of the best implementation of our generated code (generated \gemm{}) and the implementations in \cite{Benson15} (linked with Intel MKL) on single core.
More details about the experiment setup is described in Section~\ref{s:single_node}.
}
\label{tab:fmm_alg}
\end{figure}

\figref{tab:fmm_alg} summarizes a number of such algorithms that can be found in the literature that we eventually test in Section \ref{s:perf}.
We only consider $ 2 \leq \widetilde m, \widetilde k, \widetilde n \leq 6 $ and don't include arbitrary precision approximate (APA) algorithms~\cite{bini1979}, due to their questionable numerical stability.



\subsection{Kronecker Product}

If $ {X} $ and $ {Y} $  are $ m \times n $ and $ p \times q $ matrices
 with $ (i,j) $ entries denoted by $ x_{i,j} $ and $ y_{i,j} $, respectively,
then the Kronecker product \cite{KroneckerProduct}
$ {X} \otimes {Y} $
is the $ mp \times nq $ matrix given by

{
\[
{X} \otimes {Y}
=
\left(
\begin{array}{c c c}
x_{0,0} {Y} & \cdots & x_{0,n-1} {Y} \\
\vdots & \ddots  & \vdots \\
x_{m-1,0} {Y} & \cdots & x_{m-1,n-1} {Y} \\
\end{array}
\right)
\]
}
Thus,
entry $ ({X} \otimes {Y})_{p(r-1)+v,q(s-1)+w}=x_{r,s} y_{v,w} $.

\subsection{Recursive Block Indexing (Morton-like Ordering)}

\begin{figure}[bt!]
\begin{center}
{
\[
\NoShow{	{\color{red} Do we need this first matrix?}
\left(
\begin{matrix}
A_{00} &  A_{01}  & \ldots & A_{07}\\
A_{10}  & A_{11} & \ldots &A_{17}\\
\vdots & \vdots & \ddots & \vdots\\
A_{70}  &   A_{71}       &\ldots & A_{77}
\end{matrix}

\right) \\
\vspace{0.2in}
\Big\downarrow \\
\vspace{0.2in}
}
\left(
\begin{array}{c|c}
\begin{array}{c;{6pt/2pt}c}
\begin{array}{c;{2pt/2pt}c}
0 & 1 \\ \hdashline[2pt/2pt]
2 & 3
\end{array} &
\begin{array}{c;{2pt/2pt}c}
4 & 5 \\ \hdashline[2pt/2pt]
6 & 7
\end{array} \\ \hdashline[6pt/2pt]
\begin{array}{c;{2pt/2pt}c}
8 & 9 \\ \hdashline[2pt/2pt]
10 & 11
\end{array} &
\begin{array}{c;{2pt/2pt}c}
12 & 13 \\ \hdashline[2pt/2pt]
14 & 15
\end{array}
\end{array}
&
\begin{array}{c;{6pt/2pt}c}
\begin{array}{c;{2pt/2pt}c}
16 & 17 \\ \hdashline[2pt/2pt]
18 & 19
\end{array} &
\begin{array}{c;{2pt/2pt}c}
20 & 21 \\ \hdashline[2pt/2pt]
22 & 23
\end{array} \\ \hdashline[6pt/2pt]
\begin{array}{c;{2pt/2pt}c}
24 & 25 \\ \hdashline[2pt/2pt]
26 & 27
\end{array} &
\begin{array}{c;{2pt/2pt}c}
28 & 29 \\ \hdashline[2pt/2pt]
30 & 31
\end{array}
\end{array}
\\ \hline
\begin{array}{c;{6pt/2pt}c}
\begin{array}{c;{2pt/2pt}c}
32 & 33 \\ \hdashline[2pt/2pt]
34 & 35
\end{array} &
\begin{array}{c;{2pt/2pt}c}
36 & 37 \\ \hdashline[2pt/2pt]
38 & 39
\end{array} \\ \hdashline[6pt/2pt]
\begin{array}{c;{2pt/2pt}c}
40 & 41 \\ \hdashline[2pt/2pt]
42 & 43
\end{array} &
\begin{array}{c;{2pt/2pt}c}
44 & 45 \\ \hdashline[2pt/2pt]
46 & 47
\end{array}
\end{array}
&
\begin{array}{c;{6pt/2pt}c}
\begin{array}{c;{2pt/2pt}c}
48 & 49 \\ \hdashline[2pt/2pt]
50 & 51
\end{array} &
\begin{array}{c;{2pt/2pt}c}
52 & 53 \\ \hdashline[2pt/2pt]
54 & 55
\end{array} \\ \hdashline[6pt/2pt]
\begin{array}{c;{2pt/2pt}c}
56 & 57 \\ \hdashline[2pt/2pt]
58 & 59
\end{array} &
\begin{array}{c;{2pt/2pt}c}
60 & 61 \\ \hdashline[2pt/2pt]
62 & 63
\end{array}
\end{array}
\end{array}
\right)
\]
}
\end{center}
\caption{
	Illustration of recursive block storage indexing (Morton-like ordering) \cite{RecursiveBlock} on $ m \times k $ matrix $A$ where the partition dimensions $\widetilde {m}=\widetilde {k}=2$ for three-level recursions.
}
\label{fig:recursive_block_storage}
\end{figure}

An example of recursive block storage indexing (Morton-like ordering) \cite{RecursiveBlock} is given in \figref{fig:recursive_block_storage}.
In this example, $A$ undergoes three levels of recursive splitting, and each submatrix of $ A $ is indexed in row major form.
By indexing $A$, $B$, and $C$ in this manner, data locality is maintained when operations are performed on their respective submatrices. 


\subsection{Representing two-level FMM with the Kronecker Product}

In \cite{KroneckerStrassen}, it is shown that multi-level $ \langle 2, 2, 2 \rangle $ \strassen{} can be represented as Kronecker product.
In this paper, we extend this insight to multi-level FMM, where each level can use a different choice of $ \langle \widetilde{m}, \widetilde{k}, \widetilde{n}  \rangle $.

Assume each submatrix of $ A $, $ B $, and $ C $ is partitioned with another level of $ \langle \widetilde{m'}, \widetilde{k'}, \widetilde{n'} \rangle $ FMM algorithm with the coefficients $ \llbracket U', V', W' \rrbracket $, and $ A_i $, $ B_j $, $ C_p $ are the submatrices of $ A $, $ B $ and $ C $, with a single index in two-level recursive block storage indexing. Then it can be verified that $ C := C + A B $ is computed by,

%
\text{for } $ r = 0,..., R \cdot R' - 1 $,
{
\[
\begin{array}{l}
M_{r} :=
\left(\sum\limits_{i=0}^{ {\widetilde m}{\widetilde k} \cdot {\widetilde m'}{\widetilde k'}  - 1} { ({U} \otimes {U'})_{i,r} {A}_{i} } \right) \times
\left( \sum\limits_{j=0}^{ {\widetilde k}{\widetilde n} \cdot {\widetilde k'}{\widetilde n'} - 1} { ({V} \otimes {V'})_{j,r} {B}_{j} } \right)
;\\
{C}_{p} +=
({W} \otimes {W'})_{p,r}
{M}_{r}
( p = 0,..., {\widetilde m}{\widetilde n} \cdot {\widetilde m'}{\widetilde n'} - 1 )
\end{array}
\]
}
where $ \otimes $ represents Kronecker Product.
Note that
$ {U} \otimes {U'} $, $ {V} \otimes {V'} $, and $ {W} \otimes {W'} $ are
$ ( {\widetilde m}{\widetilde k} \cdot {\widetilde m'}{\widetilde k'} ) \times ( R \cdot R' ) $,
$ ( {\widetilde k}{\widetilde n} \cdot {\widetilde k'}{\widetilde n'} ) \times ( R \cdot R' ) $,
$ ( {\widetilde m}{\widetilde n} \cdot {\widetilde m'}{\widetilde n'} ) \times ( R \cdot R' ) $ matrices, respectively.

The set of coefficients of a two-level $ \langle \widetilde{m}, \widetilde{k}, \widetilde{n} \rangle $ and $ \langle \widetilde{m'}, \widetilde{k'}, \widetilde{n'} \rangle $  FMM algorithm can be denoted as
$ \llbracket {U} \otimes {U'}, {V} \otimes {V'}, {W} \otimes {W'} \rrbracket $.

For example, the two-level \strassen{} is represented by the coefficients
$ \llbracket {U} \otimes {U}, {V} \otimes {V}, {W} \otimes {W} \rrbracket $
where $ \llbracket {U}, {V}, {W} \rrbracket $ are the one-level \strassen{} coefficients given in (\ref{eqn:stra_coeffs}). 

%

\subsection{Additional levels of FMM}

Comparing one-level and two-level FMM, the same skeleton pattern emerges. The formula for defining $ L $-level FMM is given by,

\text{for } $ r = 0,..., \prod\nolimits_{l=0}^{ L - 1 }{R_l} - 1 $,
\begin{equation}
{
\begin{array}{@{}l}
M_{r} :=
\left(\sum\limits_{i=0}^{ \prod\limits_{l=0}^{L-1}{{\widetilde m_l}{\widetilde k_l}}  - 1} { (\bigotimes\limits_{l=0}^{L-1}{U_l})_{i,r} {A}_{i} } \right) \times
\left( \sum\limits_{j=0}^{ \prod\limits_{l=0}^{L-1}{{\widetilde k_l}{\widetilde n_l}}  - 1} { (\bigotimes\limits_{l=0}^{L-1}{V_l})_{j,r} {B}_{j} } \right);\\
{C}_{p} +=
(\bigotimes\limits_{l=0}^{L-1}{W_l})_{p,r}
{M}_{r}
( p = 0,..., \prod\nolimits_{l=0}^{L-1}{{\widetilde m_l}{\widetilde n_l}} - 1 )
\end{array}
}
\label{eqn:multifmm}
\end{equation}

The set of coefficients of an $ L $-level $ \langle \widetilde{m_l}, \widetilde{k_l}, \widetilde{n_l} \rangle $ ($l=0, 1,..., L-1 $) FMM algorithm can be denoted as
$ \llbracket \bigotimes\nolimits_{l=0}^{L-1}{U_l}, \bigotimes\nolimits_{l=0}^{L-1}{V_l}, \bigotimes\nolimits_{l=0}^{L-1}{W_l} \rrbracket $.



\section{Implementation and Analysis} \label{s:reload} 

\FromTo{
By this way, we represent FMM iteratively instead of recursively, which potentially avoids some extra function invocation overhead.
We will generate the skeleton code in our code generation stage, which further reduce the run time overhead.
Basically, each skeleton formula can be represented the same operation (\ref{e:fmm_1}).
The calculations of Kronecker Product and recursive block storage indexing transformations happen during the code generation stage.

We now give the details of our code generation method for families of practical fast matrix multiplication algorithms.
Next, we also present a performance model for comparing 200+ implementation variations automatically generated by our code generator.
Finally, we illustrate the parallel scheme for different implementations and various architectures, especially important for many-core architectures like KNL.
}
{
The last section shows that families of one-level FMM algorithms
can be specified by $ \langle \widetilde m, \widetilde k, \widetilde n \rangle $ and $ \llbracket U, V, W \rrbracket $.  It also shows how the Kronecker product can be used to generate multi-level FMM algorithms that are iterative rather than recursive.  In this section, we discuss a 
code generator
that takes as input $ \langle \widetilde m, \widetilde k, \widetilde n \rangle $ and $ \llbracket U, V, W \rrbracket $ and as output generates implementations that build upon the primitives that combine taking linear combinations of matrices with the packing routines and/or micro-kernels that underlie BLIS.
The code generator also provides a model of cost for each implementation that can then be used to choose the best FMM for a matrix of given size and shape. This code generator can thus generate code for arbitrary levels of FMM that can use different FMM choices at each level.  In this way, we have generated and compared more than 200 FMM algorithms.
}





\subsection{Code generation}


Our code generator generates various implementations of FMM,
based on the coefficient representation
$ \llbracket U, V, W \rrbracket $,
levels of recursion,
and packing routine/micro-kernel incorporation specifications.

There are two stages for our code generator: generating the skeleton framework, and
generating the typical operations given in (\ref{e:fmm_1}).


\subsubsection*{Generating the skeleton framework}
During this stage, the code generator
\begin{itemize}
\item Computes the Kronecker Product of the coefficient matrices $ \llbracket U_l, V_l, W_l \rrbracket $ in each level $ l $
to get the new coefficients
$ \llbracket \bigotimes\nolimits_{l=0}^{L-1}{U_l}, \bigotimes\nolimits_{l=0}^{L-1}{V_l}, \bigotimes\nolimits_{l=0}^{L-1}{W_l} \rrbracket $.

\item Generates the matrix partition code by conceptual recursive block storage indexing with
$ \langle \widetilde{m_l}, \widetilde{k_l}, \widetilde{n_l} \rangle $ partition dimensions for each level.



\item For the general cases where one or more dimensions are not multiples of corresponding $ \prod\nolimits_{l=0}^{L-1}{\widetilde {m_l}} $, $ \prod\nolimits_{l=0}^{L-1}{\widetilde {k_l}} $, $ \prod\nolimits_{l=0}^{L-1}{\widetilde {n_l}} $, it generates dynamic peeling \cite{StrassenDynamicPeeling} code to handle the remaining ``fringes'' after invoking FMM, which requires no additional memory.

\end{itemize}


\subsubsection*{Generating the typical operations}


To generate the code for the typical operations in (\ref{e:fmm_1}),
the code generator

\begin{itemize}
\item Generates  packing routines (written in {\tt C}), that sum a list of submatrices of $ A $ integrated into the packing routine, yielding $ \widetilde A_i $,
and similarly sum a list of submatrices of $ B $ integrated into the packing routine, yielding $ \widetilde B_p $, extending what is illustrated in \figref{fig:side_by_side}.

\NoShow{\item Generates the specialized ({\tt assembly}-coded) micro-kernel that updates multiple submatrices of $ C $ with a submatrix of $ M_r $ that resides in the registers.}
\item Assembles a specialized micro-kernel comprised of a hand-coded \gemm{} kernel and automatically generated updates to multiple submatrices of $ C $.

\NoShow{We use templates to generate the {\tt assembly} code for the main loop in the micro-kernel.
After the computation for $ M_r $ is finished,
our code generator allocates the registers for updating different $ C_p $ with an efficient round robin policy to reduce instruction dependencies
and improve the performance.
}
\end{itemize}

\subsubsection*{Further variations}
In~\cite{Strassen:SC16}, a number of variations on the theme illustrated in \figref{fig:side_by_side}~(right) are discussed:
\NoShow{The members of our generated families of FMM implementations differ by
which partition dimension $ \langle \widetilde m, \widetilde k, \widetilde n \rangle $ in each level,
how many levels of FMM they incorporate and
which of the above described generated packing routines/micro-kernel they use:}
\begin{itemize}
	\item
	\XXXFMM{}: \FromTo{The code generator generates a}{A} classical implementation with temporary buffers for storing the sum of $ A $, $ B $, and the intermediate matrix product $ M_r $.
	\item
	\ABXFMM{}: \FromTo{The code generator generates t}{T}he packing routines
	\FromTo{which}{} incorporate the summation of submatrices of $ A $, $ B $ into the packing of buffers $ \widetilde A_i $ and $ \widetilde B_p $ but \FromTo{creates}{} explicit temporary buffers for matrices $ M_r $ \FromTo{}{are used}.
	\item
	\ABCFMM{}: \FromTo{The code generator generates the packing routines which incorporate the summation of submatrices of $ A $, $ B $ into the packing of buffers $ \widetilde A_i $ and $ \widetilde B_p $
{\em and}}{} \ABXFMM, but with a specialized micro-kernel that incorporates addition of $ M_r $ to multiple submatrices of $ C $.
\end{itemize}
Incorporating the generation of these variations into the code generator
yields over 200 FMM implementations.

\subsection{Performance model}

\begin{figure}[!t]
\centering
{
\setlength{\tabcolsep}{2pt}
  \begin{tabular}{c | l }
  \whline
  \multirow{2}{*}{$ \tau_a $} & Time (in sec.) of one \underline{a}rithmetic (floating point), \\
  & operation, reciprocal of theoretical peak \GFLOPS{}. \\
\hline
  \multirow{2}{*}{$ \tau_b $} & (\underline{B}andwidth) Amortized time (in sec.) of 8 Bytes \\
  & contigous data movement from DRAM to cache. \\
\hline
  $ T $     & Total execution time (in sec.). \\
\hline
  $ T_a $   & Time for \underline{a}rithmetic operations (in sec. ). \\
\hline
  $ T_m $   & Time for \underline{m}emory operations (in sec. ). \\
\hline
  $T_{a}^{\times}$ & $ T_a $ for submatrix multiplications. \\
\hline
  $T_{a}^{A_{+}}$, $T_{a}^{B_{+}}$, $T_{a}^{C_{+}}$ & $ T_a $ for extra submatrix additions. \\
\hline
  $T_{m}^{A_{\times}}$, $T_{m}^{B_{\times}}$ & $ T_m $ for reading submatrices in packing routines (Fig. \ref{fig:side_by_side}).\\
\hline
  $T_{m}^{{\widetilde A}_{\times}}$,$T_{m}^{{\widetilde B}_{\times}}$  & $ T_m $ for writing submatrices in packing routines (Fig. \ref{fig:side_by_side}).\\
\hline
  \multirow{2}{*}{$T_{m}^{C_{\times}}$} & $ T_m $ for reading \emph{and} writing submatrices in  micro-kernel\\
   & (Fig. \ref{fig:side_by_side}).\\
\hline
  \multirow{2}{*}{$T_{m}^{A_{+}}$, $T_{m}^{B_{+}}$, $T_{m}^{C_{+}}$} &
$ T_m $ for reading \emph{or} writing submatrices, related to the \\
   & temporary buffer as part of \XXXFMM{} and \ABXFMM{}. \\
\hline
  $ nnz(X) $ & non-zero entry number in matrix or vector $ X $. \\
  \whline
  \end{tabular}
}
\caption{Notation table for performance model.}
\label{tab:notation}
\end{figure}

\begin{figure}[p]
\centering
{
  \begin{tabular}{l | l cr r }
  \whline
  \mycircle{1} & $\text{\emph{Effective} GFLOPS} = 2\cdot m\cdot n\cdot k / T \cdot 10 ^{-9}$ \\
\hline
  \mycircle{2} & $T=T_{a}+T_{m}$ \\
\hline
  {\mycircle{3}} & $T_{a} = N_{a}^{\times} \cdot T_{a}^{\times} + N_{a}^{A_{+}} \cdot T_{a}^{A_{+}} + N_{a}^{B_{+}} \cdot T_{a}^{B_{+}} + N_{a}^{C_{+}} \cdot T_{a}^{C_{+}} $  \\
\hline
  {\mycircle{4}} & $T_{m} = N_{m}^{A_{\times}} \cdot T_{m}^{A_{\times}} + N_{m}^{B_{\times}} \cdot T_{m}^{B_{\times}} + N_{m}^{C_{\times}} \cdot T_{m}^{C_{\times}} + N_{m}^{A_{+}} \cdot T_{m}^{A_{+}} + N_{m}^{B_{+}} \cdot T_{m}^{B_{+}} + N_{m}^{C_{+}} \cdot T_{m}^{C_{+}}$ \\
  \whline
  \end{tabular}
}

\vspace{0.2in}

\centering
{
  \begin{tabular}{l| ccr r }
  \whline
                                        & type
                                        & $\tau$
                                        & \gemm{}
                                        & $L$-level  \\
  \hline
  $T_{a}^{\times}$                      & -
                                        & $\tau_{a}$
                                        & $ 2 m  n  k $
                                        & $ 2 \frac{m}{\DivisorML{}}\frac{n}{\DivisorNL}\frac{k}{\DivisorKL}$ \\
  $T_{a}^{A_{+}}$                       & -
                                        & $\tau_{a}$
                                        & -
                                        & $ 2 \frac{m}{\DivisorML{}}\frac{k}{\DivisorKL} $ \\
  $T_{a}^{B_{+}}$                       & -
                                        & $\tau_{a}$
                                        & -
                                        & $ 2 \frac{k}{\DivisorKL{}}\frac{n}{\DivisorNL{}} $ \\
  $T_{a}^{C_{+}}$                       & -
                                        & $\tau_{a}$
                                        & -
                                        & $ 2 \frac{m}{\DivisorML{}}\frac{n}{\DivisorNL{}} $ \\
  \hline
  $T_{m}^{A_{\times}}$                  & \texttt{r}
                                        & $\tau_{b}$
                                        & $mk \lceil \frac{n}{n_c} \rceil$
                                        & $\frac{m}{ \DivisorML{}} \frac{k}{ \DivisorKL{} } \lceil \frac{n/\DivisorNL{}}{n_c} \rceil$ \\
  $T_{m}^{{\widetilde A}_{\times}}$     & \texttt{w}
                                        & $\tau_{b}$
                                        & $mk \lceil \frac{n}{n_c} \rceil$
                                        & $\frac{m}{\DivisorML{}} \frac{k}{ \DivisorKL{}} \lceil \frac{n/\DivisorNL{}}{n_c} \rceil$ \\
  $T_{m}^{B_{\times}}$                  & \texttt{r}
                                        & $\tau_{b}$
                                        & $nk$
                                        & $\frac{n}{\DivisorNL{}} \frac{k}{ \DivisorKL{}}$ \\
  $T_{m}^{{\widetilde B}_{\times}}$       & \texttt{w}
                                        & $\tau_{b}$
                                        & $nk$
                                        & $\frac{n}{ \DivisorNL{}} \frac{k}{\DivisorKL{}}$ \\
  $T_{m}^{C_{\times}}$ \fromto{(*)}{}   & \texttt{r/w}
                                        & $\tau_{b}$
                                        & $2\lambda mn\lceil\frac{k}{k_c}\rceil$
                                        & $2\lambda \frac{m}{\DivisorML{}}\frac{n}{\DivisorNL{}}\lceil\frac{k/\DivisorKL{}}{k_c}\rceil $ \\
  \hline
  $T_{m}^{A_{+}}$                       & \texttt{r/w}
                                        & $\tau_{b}$
                                        & $mk$
                                        & $\frac{m}{\DivisorML{}} \frac{k}{\DivisorKL{}}$ \\
  $T_{m}^{B_{+}}$                       & \texttt{r/w}
                                        & $\tau_{b}$
                                        & $nk$
                                        & $\frac{n}{\DivisorNL{}} \frac{k}{\DivisorKL{}}$ \\
  $T_{m}^{C_{+}}$                       & \texttt{r/w}
                                        & $\tau_{b}$
                                        & $mn$
                                        & $\frac{m}{\DivisorML{}} \frac{n}{\DivisorNL{}}$ \\
  \whline
  \end{tabular}
}

\vspace{0.2in}


\centering
{
\setlength{\tabcolsep}{3pt}
  \begin{tabular}{c | c | c |c |c }
  \whline
                                        & \multirow{2}{*}{\gemm{}}    &   \multicolumn{3}{c} { $ L $-level }   \\ \cline{3-5}
                                        &                              &      ABC
                                                                       &      AB
                                                                       &      Naive \\
  \hline
  $N_{a}^{\times}$                      & 1
                                        & $ \PRODRL{} $
                                        & $ \PRODRL{} $
                                        & $ \PRODRL{} $ \\
  $N_{a}^{A_{+}}$                       & -
                                        & $ nnz(\OTIMESU{}) \text{-} \PRODRL{} $
                                        & $ nnz(\OTIMESU{}) \text{-} \PRODRL{} $
                                        & $ nnz(\OTIMESU{}) \text{-} \PRODRL{} $ \\
  $N_{a}^{B_{+}}$                       & -
                                        & $ nnz(\OTIMESV{}) \text{-} \PRODRL{} $
                                        & $ nnz(\OTIMESV{}) \text{-} \PRODRL{} $
                                        & $ nnz(\OTIMESV{}) \text{-} \PRODRL{} $ \\
  $N_{a}^{C_{+}}$                       & -
                                        & $ nnz(\OTIMESW{}) $
                                        & $ nnz(\OTIMESW{}) $
                                        & $ nnz(\OTIMESW{}) $ \\
  \hline
  $N_{m}^{A_{\times}}$                  & 1
                                        & $ nnz(\OTIMESU{}) $
                                        & $ nnz(\OTIMESU{}) $
                                        & $ \PRODRL{} $ \\
  $N_{m}^{{\widetilde A}_{\times}}$     & -
                                        & -
                                        & -
                                        & - \\
  $N_{m}^{B_{\times}}$                  & $ 1  $
                                        & $ nnz(\OTIMESV{}) $
                                        & $ nnz(\OTIMESV{}) $
                                        & $ \PRODRL{} $ \\
  $N_{m}^{{\widetilde B}_{\times}}$     & -
                                        & -
                                        & -
                                        & - \\
  $N_{m}^{C_{\times}}$                  & 1
                                        & $ nnz(\OTIMESW{}) $
                                        & $ \PRODRL{} $
                                        & $ \PRODRL{} $ \\
  \hline
  $N_{m}^{A_{+}}$                       & -
                                        & -
                                        & -
                                        & $ nnz(\OTIMESU{})\text{+}\PRODRL{} $ \\
  $N_{m}^{B_{+}}$                       & -
                                        & -
                                        & -
                                        & $ nnz(\OTIMESV{})\text{+}\PRODRL{} $ \\
  $N_{m}^{C_{+}}$                       & -
                                        & -
                                        & $ 3 nnz(\OTIMESW{}) $
                                        & $ 3 nnz(\OTIMESW{}) $ \\
  \whline
  \end{tabular}

}

\caption{
The top table shows
the equations for computing the execution time $ T $ and \emph{Effective} GFLOPS in our performance model.
The middle table shows the various components of arithmetic and memory operations for BLAS \gemm{} and various implementations of \FMM{}.
The time shown in the first column for \gemm{} and $L$-level \FMM{}
can be computed separately by multiplying the parameter in $\tau$ column with
the arithmetic/memory operation number in the corresponding entries.
The bottom table shows
the coefficient $N_a^X$/$N_m^X$ mapping table for computing $T_a$/$T_m$ in the performance model.
Here $ \DivisorML = \prod\nolimits_{l=0}^{L-1}{\widetilde {m_l}} $,
$ \DivisorKL = \prod\nolimits_{l=0}^{L-1}{\widetilde {k_l}} $,
$ \DivisorNL = \prod\nolimits_{l=0}^{L-1}{\widetilde {n_l}} $,
$ \OTIMESU   = \bigotimes\nolimits_{l=0}^{L-1}{U_l} $,
$ \OTIMESV   = \bigotimes\nolimits_{l=0}^{L-1}{V_l} $,
$ \OTIMESW   = \bigotimes\nolimits_{l=0}^{L-1}{W_l} $,
$ \PRODRL    = \prod\nolimits_{l=0}^{L-1}{ R_l } $.
}
\label{tab:perfmodel}
\end{figure}

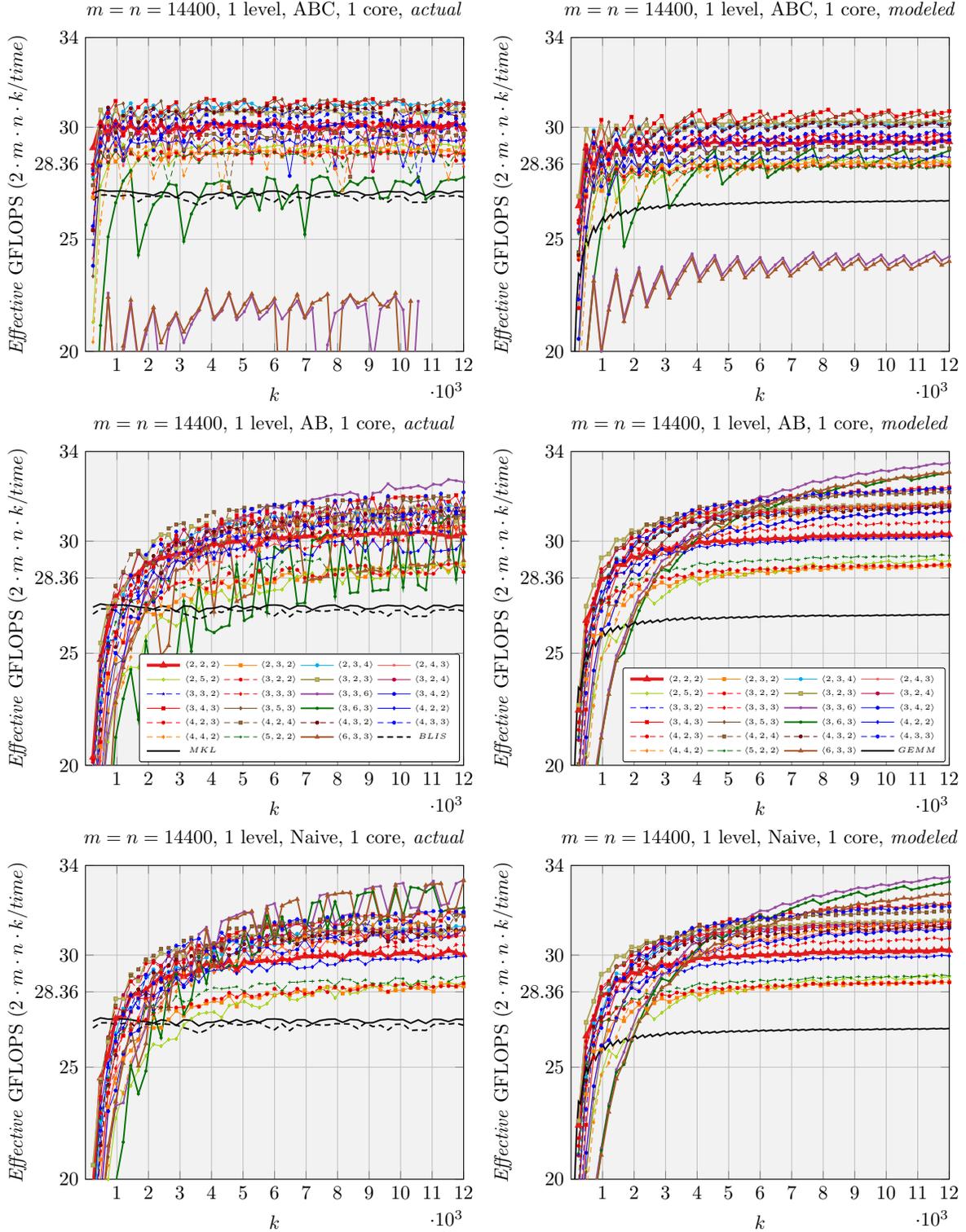
\begin{figure*}[htp!]
\center
\begin{tikzpicture}[scale=0.9]
\begin{axis}[
    title={$m = n = 14400$, 1 level, ABC, 1 core, \emph{actual}},
    xlabel={ $k$ },
    ylabel={\emph{Effective} GFLOPS ($ 2 \cdot  m \cdot  n \cdot k / time $)},
    xmin=0,
    xmax=12000,
    ymin=20,
    ymax=34,
    xtick={1000,2000,3000,4000,5000,6000,7000,8000,9000,10000,11000,12000},
    ytick={5, 10, 15, 20, 25, 28.36, 30, 34},
    scaled x ticks=false,
    scaled x ticks=base 10:-3,
    grid=major,
    axis background/.style={fill=lightgray!20},
    mark size=0.8pt,
    cycle list name=jianyucolor,
    restrict y to domain=1:inf,
    ]
\addplot table[x=dim,y=222,col sep=comma] {plotdata/rankk_1level_abc_1core.csv};
\addplot table[x=dim,y=232,col sep=comma] {plotdata/rankk_1level_abc_1core.csv};
\addplot table[x=dim,y=234,col sep=comma] {plotdata/rankk_1level_abc_1core.csv};
\addplot table[x=dim,y=243,col sep=comma] {plotdata/rankk_1level_abc_1core.csv};
\addplot table[x=dim,y=252,col sep=comma] {plotdata/rankk_1level_abc_1core.csv};
\addplot table[x=dim,y=322,col sep=comma] {plotdata/rankk_1level_abc_1core.csv};
\addplot table[x=dim,y=323,col sep=comma] {plotdata/rankk_1level_abc_1core.csv};
\addplot table[x=dim,y=324,col sep=comma] {plotdata/rankk_1level_abc_1core.csv};
\addplot table[x=dim,y=332,col sep=comma] {plotdata/rankk_1level_abc_1core.csv};
\addplot table[x=dim,y=333,col sep=comma] {plotdata/rankk_1level_abc_1core.csv};
\addplot table[x=dim,y=336,col sep=comma] {plotdata/rankk_1level_abc_1core.csv};
\addplot table[x=dim,y=342,col sep=comma] {plotdata/rankk_1level_abc_1core.csv};
\addplot table[x=dim,y=343,col sep=comma] {plotdata/rankk_1level_abc_1core.csv};
\addplot table[x=dim,y=353,col sep=comma] {plotdata/rankk_1level_abc_1core.csv};
\addplot table[x=dim,y=363,col sep=comma] {plotdata/rankk_1level_abc_1core.csv};
\addplot table[x=dim,y=422,col sep=comma] {plotdata/rankk_1level_abc_1core.csv};
\addplot table[x=dim,y=423,col sep=comma] {plotdata/rankk_1level_abc_1core.csv};
\addplot table[x=dim,y=424,col sep=comma] {plotdata/rankk_1level_abc_1core.csv};
\addplot table[x=dim,y=432,col sep=comma] {plotdata/rankk_1level_abc_1core.csv};
\addplot table[x=dim,y=433,col sep=comma] {plotdata/rankk_1level_abc_1core.csv};
\addplot table[x=dim,y=442,col sep=comma] {plotdata/rankk_1level_abc_1core.csv};
\addplot table[x=dim,y=522,col sep=comma] {plotdata/rankk_1level_abc_1core.csv};
\addplot table[x=dim,y=633,col sep=comma] {plotdata/rankk_1level_abc_1core.csv};
\addplot table[x=dim,y=blis,col sep=comma] {plotdata/rankk_1level_abc_1core.csv};
\addplot table[x=dim,y=mkl,col sep=comma] {plotdata/rankk_1level_abc_1core.csv};
\end{axis}
\end{tikzpicture}
\begin{tikzpicture}[scale=0.9]
\begin{axis}[
    title={$m = n = 14400$, 1 level, ABC, 1 core, \emph{modeled}},
    xlabel={ $k$ },
    ylabel={\emph{Effective} GFLOPS ($ 2 \cdot  m \cdot  n \cdot k / time $)},
    xmin=0,
    xmax=12000,
    ymin=20,
    ymax=34,
    xtick={1000,2000,3000,4000,5000,6000,7000,8000,9000,10000,11000,12000},
    ytick={5, 10, 15, 20, 25, 28.36, 30, 34},
    scaled x ticks=false,
    scaled x ticks=base 10:-3,
    grid=major,
    axis background/.style={fill=lightgray!20},
    mark size=0.8pt,
    cycle list name=jianyu0color,
    ]
\addplot table[x=dim,y=222,col sep=comma] {plotdata/model_rankk_1level_abc_1core.csv};
\addplot table[x=dim,y=232,col sep=comma] {plotdata/model_rankk_1level_abc_1core.csv};
\addplot table[x=dim,y=234,col sep=comma] {plotdata/model_rankk_1level_abc_1core.csv};
\addplot table[x=dim,y=243,col sep=comma] {plotdata/model_rankk_1level_abc_1core.csv};
\addplot table[x=dim,y=252,col sep=comma] {plotdata/model_rankk_1level_abc_1core.csv};
\addplot table[x=dim,y=322,col sep=comma] {plotdata/model_rankk_1level_abc_1core.csv};
\addplot table[x=dim,y=323,col sep=comma] {plotdata/model_rankk_1level_abc_1core.csv};
\addplot table[x=dim,y=324,col sep=comma] {plotdata/model_rankk_1level_abc_1core.csv};
\addplot table[x=dim,y=332,col sep=comma] {plotdata/model_rankk_1level_abc_1core.csv};
\addplot table[x=dim,y=333,col sep=comma] {plotdata/model_rankk_1level_abc_1core.csv};
\addplot table[x=dim,y=336,col sep=comma] {plotdata/model_rankk_1level_abc_1core.csv};
\addplot table[x=dim,y=342,col sep=comma] {plotdata/model_rankk_1level_abc_1core.csv};
\addplot table[x=dim,y=343,col sep=comma] {plotdata/model_rankk_1level_abc_1core.csv};
\addplot table[x=dim,y=353,col sep=comma] {plotdata/model_rankk_1level_abc_1core.csv};
\addplot table[x=dim,y=363,col sep=comma] {plotdata/model_rankk_1level_abc_1core.csv};
\addplot table[x=dim,y=422,col sep=comma] {plotdata/model_rankk_1level_abc_1core.csv};
\addplot table[x=dim,y=423,col sep=comma] {plotdata/model_rankk_1level_abc_1core.csv};
\addplot table[x=dim,y=424,col sep=comma] {plotdata/model_rankk_1level_abc_1core.csv};
\addplot table[x=dim,y=432,col sep=comma] {plotdata/model_rankk_1level_abc_1core.csv};
\addplot table[x=dim,y=433,col sep=comma] {plotdata/model_rankk_1level_abc_1core.csv};
\addplot table[x=dim,y=442,col sep=comma] {plotdata/model_rankk_1level_abc_1core.csv};
\addplot table[x=dim,y=522,col sep=comma] {plotdata/model_rankk_1level_abc_1core.csv};
\addplot table[x=dim,y=633,col sep=comma] {plotdata/model_rankk_1level_abc_1core.csv};
\addplot table[x=dim,y=gemm,col sep=comma] {plotdata/model_rankk_gemm_1core.csv};
\end{axis}
\end{tikzpicture}
\begin{tikzpicture}[scale=0.9]
\begin{axis}[
    title={$m = n = 14400$, 1 level, AB, 1 core, \emph{actual}},
    xlabel={ $k$ },
    ylabel={\emph{Effective} GFLOPS ($ 2 \cdot  m \cdot  n \cdot k / time $)},
    xmin=0,
    xmax=12000,
    ymin=20,
    ymax=34,
    xtick={1000,2000,3000,4000,5000,6000,7000,8000,9000,10000,11000,12000},
    ytick={5, 10, 15, 20, 25, 28.36, 30, 34},
    scaled x ticks=false,
    scaled x ticks=base 10:-3,
    grid=major,
    axis background/.style={fill=lightgray!20},
    mark size=0.8pt,
    cycle list name=jianyucolor,
    restrict y to domain=1:inf,
    legend style={
        at={(0.99,0.01)},
        anchor=south east,
        legend columns=4,
        font=\tiny,
        rounded corners=1pt,
        nodes={scale=0.8, transform shape},
    },
    legend entries = {
        $ \langle 2, 2, 2 \rangle $\\
        $ \langle 2, 3, 2 \rangle $\\ 
        $ \langle 2, 3, 4 \rangle $\\ 
        $ \langle 2, 4, 3 \rangle $\\ 
        $ \langle 2, 5, 2 \rangle $\\ 
        $ \langle 3, 2, 2 \rangle $\\ 
        $ \langle 3, 2, 3 \rangle $\\ 
        $ \langle 3, 2, 4 \rangle $\\ 
        $ \langle 3, 3, 2 \rangle $\\ 
        $ \langle 3, 3, 3 \rangle $\\ 
        $ \langle 3, 3, 6 \rangle $\\ 
        $ \langle 3, 4, 2 \rangle $\\ 
        $ \langle 3, 4, 3 \rangle $\\ 
        $ \langle 3, 5, 3 \rangle $\\ 
        $ \langle 3, 6, 3 \rangle $\\ 
        $ \langle 4, 2, 2 \rangle $\\ 
        $ \langle 4, 2, 3 \rangle $\\ 
        $ \langle 4, 2, 4 \rangle $\\ 
        $ \langle 4, 3, 2 \rangle $\\ 
        $ \langle 4, 3, 3 \rangle $\\ 
        $ \langle 4, 4, 2 \rangle $\\ 
        $ \langle 5, 2, 2 \rangle $\\ 
        $ \langle 6, 3, 3 \rangle $\\
        $ BLIS $\\
        $ MKL $\\
    },
    ]
\addplot table[x=dim,y=222,col sep=comma] {plotdata/rankk_1level_ab_1core.csv};
\addplot table[x=dim,y=232,col sep=comma] {plotdata/rankk_1level_ab_1core.csv};
\addplot table[x=dim,y=234,col sep=comma] {plotdata/rankk_1level_ab_1core.csv};
\addplot table[x=dim,y=243,col sep=comma] {plotdata/rankk_1level_ab_1core.csv};
\addplot table[x=dim,y=252,col sep=comma] {plotdata/rankk_1level_ab_1core.csv};
\addplot table[x=dim,y=322,col sep=comma] {plotdata/rankk_1level_ab_1core.csv};
\addplot table[x=dim,y=323,col sep=comma] {plotdata/rankk_1level_ab_1core.csv};
\addplot table[x=dim,y=324,col sep=comma] {plotdata/rankk_1level_ab_1core.csv};
\addplot table[x=dim,y=332,col sep=comma] {plotdata/rankk_1level_ab_1core.csv};
\addplot table[x=dim,y=333,col sep=comma] {plotdata/rankk_1level_ab_1core.csv};
\addplot table[x=dim,y=336,col sep=comma] {plotdata/rankk_1level_ab_1core.csv};
\addplot table[x=dim,y=342,col sep=comma] {plotdata/rankk_1level_ab_1core.csv};
\addplot table[x=dim,y=343,col sep=comma] {plotdata/rankk_1level_ab_1core.csv};
\addplot table[x=dim,y=353,col sep=comma] {plotdata/rankk_1level_ab_1core.csv};
\addplot table[x=dim,y=363,col sep=comma] {plotdata/rankk_1level_ab_1core.csv};
\addplot table[x=dim,y=422,col sep=comma] {plotdata/rankk_1level_ab_1core.csv};
\addplot table[x=dim,y=423,col sep=comma] {plotdata/rankk_1level_ab_1core.csv};
\addplot table[x=dim,y=424,col sep=comma] {plotdata/rankk_1level_ab_1core.csv};
\addplot table[x=dim,y=432,col sep=comma] {plotdata/rankk_1level_ab_1core.csv};
\addplot table[x=dim,y=433,col sep=comma] {plotdata/rankk_1level_ab_1core.csv};
\addplot table[x=dim,y=442,col sep=comma] {plotdata/rankk_1level_ab_1core.csv};
\addplot table[x=dim,y=522,col sep=comma] {plotdata/rankk_1level_ab_1core.csv};
\addplot table[x=dim,y=633,col sep=comma] {plotdata/rankk_1level_ab_1core.csv};
\addplot table[x=dim,y=blis,col sep=comma] {plotdata/rankk_1level_ab_1core.csv};
\addplot table[x=dim,y=mkl,col sep=comma] {plotdata/rankk_1level_ab_1core.csv};
\end{axis}
\end{tikzpicture}
\begin{tikzpicture}[scale=0.9]
\begin{axis}[
    title={$m = n = 14400$, 1 level, AB, 1 core, \emph{modeled}},
    xlabel={ $k$ },
    ylabel={\emph{Effective} GFLOPS ($ 2 \cdot  m \cdot  n \cdot k / time $)},
    xmin=0,
    xmax=12000,
    ymin=20,
    ymax=34,
    xtick={1000,2000,3000,4000,5000,6000,7000,8000,9000,10000,11000,12000},
    ytick={5, 10, 15, 20, 25, 28.36, 30, 34},
    scaled x ticks=false,
    scaled x ticks=base 10:-3,
    grid=major,
    axis background/.style={fill=lightgray!20},
    mark size=0.8pt,
    cycle list name=jianyu0color,
    legend style={
        at={(0.99,0.01)},
        anchor=south east,
        legend columns=4,
        font=\tiny,
        rounded corners=1pt,
        nodes={scale=0.8, transform shape},
    },
    legend entries = {
        $ \langle 2, 2, 2 \rangle $\\
        $ \langle 2, 3, 2 \rangle $\\ 
        $ \langle 2, 3, 4 \rangle $\\ 
        $ \langle 2, 4, 3 \rangle $\\ 
        $ \langle 2, 5, 2 \rangle $\\ 
        $ \langle 3, 2, 2 \rangle $\\ 
        $ \langle 3, 2, 3 \rangle $\\ 
        $ \langle 3, 2, 4 \rangle $\\ 
        $ \langle 3, 3, 2 \rangle $\\ 
        $ \langle 3, 3, 3 \rangle $\\ 
        $ \langle 3, 3, 6 \rangle $\\ 
        $ \langle 3, 4, 2 \rangle $\\ 
        $ \langle 3, 4, 3 \rangle $\\ 
        $ \langle 3, 5, 3 \rangle $\\ 
        $ \langle 3, 6, 3 \rangle $\\ 
        $ \langle 4, 2, 2 \rangle $\\ 
        $ \langle 4, 2, 3 \rangle $\\ 
        $ \langle 4, 2, 4 \rangle $\\ 
        $ \langle 4, 3, 2 \rangle $\\ 
        $ \langle 4, 3, 3 \rangle $\\ 
        $ \langle 4, 4, 2 \rangle $\\ 
        $ \langle 5, 2, 2 \rangle $\\ 
        $ \langle 6, 3, 3 \rangle $\\
        $ GEMM     $\\
    },
    ]
\addplot table[x=dim,y=222,col sep=comma] {plotdata/model_rankk_1level_ab_1core.csv};
\addplot table[x=dim,y=232,col sep=comma] {plotdata/model_rankk_1level_ab_1core.csv};
\addplot table[x=dim,y=234,col sep=comma] {plotdata/model_rankk_1level_ab_1core.csv};
\addplot table[x=dim,y=243,col sep=comma] {plotdata/model_rankk_1level_ab_1core.csv};
\addplot table[x=dim,y=252,col sep=comma] {plotdata/model_rankk_1level_ab_1core.csv};
\addplot table[x=dim,y=322,col sep=comma] {plotdata/model_rankk_1level_ab_1core.csv};
\addplot table[x=dim,y=323,col sep=comma] {plotdata/model_rankk_1level_ab_1core.csv};
\addplot table[x=dim,y=324,col sep=comma] {plotdata/model_rankk_1level_ab_1core.csv};
\addplot table[x=dim,y=332,col sep=comma] {plotdata/model_rankk_1level_ab_1core.csv};
\addplot table[x=dim,y=333,col sep=comma] {plotdata/model_rankk_1level_ab_1core.csv};
\addplot table[x=dim,y=336,col sep=comma] {plotdata/model_rankk_1level_ab_1core.csv};
\addplot table[x=dim,y=342,col sep=comma] {plotdata/model_rankk_1level_ab_1core.csv};
\addplot table[x=dim,y=343,col sep=comma] {plotdata/model_rankk_1level_ab_1core.csv};
\addplot table[x=dim,y=353,col sep=comma] {plotdata/model_rankk_1level_ab_1core.csv};
\addplot table[x=dim,y=363,col sep=comma] {plotdata/model_rankk_1level_ab_1core.csv};
\addplot table[x=dim,y=422,col sep=comma] {plotdata/model_rankk_1level_ab_1core.csv};
\addplot table[x=dim,y=423,col sep=comma] {plotdata/model_rankk_1level_ab_1core.csv};
\addplot table[x=dim,y=424,col sep=comma] {plotdata/model_rankk_1level_ab_1core.csv};
\addplot table[x=dim,y=432,col sep=comma] {plotdata/model_rankk_1level_ab_1core.csv};
\addplot table[x=dim,y=433,col sep=comma] {plotdata/model_rankk_1level_ab_1core.csv};
\addplot table[x=dim,y=442,col sep=comma] {plotdata/model_rankk_1level_ab_1core.csv};
\addplot table[x=dim,y=522,col sep=comma] {plotdata/model_rankk_1level_ab_1core.csv};
\addplot table[x=dim,y=633,col sep=comma] {plotdata/model_rankk_1level_ab_1core.csv};
\addplot table[x=dim,y=gemm,col sep=comma] {plotdata/model_rankk_gemm_1core.csv};
\end{axis}
\end{tikzpicture}
\begin{tikzpicture}[scale=0.9]
\begin{axis}[
    title={$m = n = 14400$, 1 level, Naive, 1 core, \emph{actual}},
    xlabel={ $k$ },
    ylabel={\emph{Effective} GFLOPS ($ 2 \cdot  m \cdot  n \cdot k / time $)},
    xmin=0,
    xmax=12000,
    ymin=20,
    ymax=34,
    xtick={1000,2000,3000,4000,5000,6000,7000,8000,9000,10000,11000,12000},
    ytick={5, 10, 15, 20, 25, 28.36, 30, 34},
    scaled x ticks=false,
    scaled x ticks=base 10:-3,
    grid=major,
    axis background/.style={fill=lightgray!20},
    mark size=0.8pt,
    cycle list name=jianyucolor,
    restrict y to domain=1:inf,
    ]
\addplot table[x=dim,y=222,col sep=comma] {plotdata/rankk_1level_naive_1core.csv};
\addplot table[x=dim,y=232,col sep=comma] {plotdata/rankk_1level_naive_1core.csv};
\addplot table[x=dim,y=234,col sep=comma] {plotdata/rankk_1level_naive_1core.csv};
\addplot table[x=dim,y=243,col sep=comma] {plotdata/rankk_1level_naive_1core.csv};
\addplot table[x=dim,y=252,col sep=comma] {plotdata/rankk_1level_naive_1core.csv};
\addplot table[x=dim,y=322,col sep=comma] {plotdata/rankk_1level_naive_1core.csv};
\addplot table[x=dim,y=323,col sep=comma] {plotdata/rankk_1level_naive_1core.csv};
\addplot table[x=dim,y=324,col sep=comma] {plotdata/rankk_1level_naive_1core.csv};
\addplot table[x=dim,y=332,col sep=comma] {plotdata/rankk_1level_naive_1core.csv};
\addplot table[x=dim,y=333,col sep=comma] {plotdata/rankk_1level_naive_1core.csv};
\addplot table[x=dim,y=336,col sep=comma] {plotdata/rankk_1level_naive_1core.csv};
\addplot table[x=dim,y=342,col sep=comma] {plotdata/rankk_1level_naive_1core.csv};
\addplot table[x=dim,y=343,col sep=comma] {plotdata/rankk_1level_naive_1core.csv};
\addplot table[x=dim,y=353,col sep=comma] {plotdata/rankk_1level_naive_1core.csv};
\addplot table[x=dim,y=363,col sep=comma] {plotdata/rankk_1level_naive_1core.csv};
\addplot table[x=dim,y=422,col sep=comma] {plotdata/rankk_1level_naive_1core.csv};
\addplot table[x=dim,y=423,col sep=comma] {plotdata/rankk_1level_naive_1core.csv};
\addplot table[x=dim,y=424,col sep=comma] {plotdata/rankk_1level_naive_1core.csv};
\addplot table[x=dim,y=432,col sep=comma] {plotdata/rankk_1level_naive_1core.csv};
\addplot table[x=dim,y=433,col sep=comma] {plotdata/rankk_1level_naive_1core.csv};
\addplot table[x=dim,y=442,col sep=comma] {plotdata/rankk_1level_naive_1core.csv};
\addplot table[x=dim,y=522,col sep=comma] {plotdata/rankk_1level_naive_1core.csv};
\addplot table[x=dim,y=633,col sep=comma] {plotdata/rankk_1level_naive_1core.csv};
\addplot table[x=dim,y=blis,col sep=comma] {plotdata/rankk_1level_naive_1core.csv};
\addplot table[x=dim,y=mkl,col sep=comma] {plotdata/rankk_1level_naive_1core.csv};
\end{axis}
\end{tikzpicture}
\begin{tikzpicture}[scale=0.9]
\begin{axis}[
    title={$m = n = 14400$, 1 level, Naive, 1 core, \emph{modeled}},
    xlabel={ $k$ },
    ylabel={\emph{Effective} GFLOPS ($ 2 \cdot  m \cdot  n \cdot k / time $)},
    xmin=0,
    xmax=12000,
    ymin=20,
    ymax=34,
    xtick={1000,2000,3000,4000,5000,6000,7000,8000,9000,10000,11000,12000},
    ytick={5, 10, 15, 20, 25, 28.36, 30, 34},
    scaled x ticks=false,
    scaled x ticks=base 10:-3,
    grid=major,
    axis background/.style={fill=lightgray!20},
    mark size=0.8pt,
    cycle list name=jianyu0color,
    ]
\addplot table[x=dim,y=222,col sep=comma] {plotdata/model_rankk_1level_naive_1core.csv};
\addplot table[x=dim,y=232,col sep=comma] {plotdata/model_rankk_1level_naive_1core.csv};
\addplot table[x=dim,y=234,col sep=comma] {plotdata/model_rankk_1level_naive_1core.csv};
\addplot table[x=dim,y=243,col sep=comma] {plotdata/model_rankk_1level_naive_1core.csv};
\addplot table[x=dim,y=252,col sep=comma] {plotdata/model_rankk_1level_naive_1core.csv};
\addplot table[x=dim,y=322,col sep=comma] {plotdata/model_rankk_1level_naive_1core.csv};
\addplot table[x=dim,y=323,col sep=comma] {plotdata/model_rankk_1level_naive_1core.csv};
\addplot table[x=dim,y=324,col sep=comma] {plotdata/model_rankk_1level_naive_1core.csv};
\addplot table[x=dim,y=332,col sep=comma] {plotdata/model_rankk_1level_naive_1core.csv};
\addplot table[x=dim,y=333,col sep=comma] {plotdata/model_rankk_1level_naive_1core.csv};
\addplot table[x=dim,y=336,col sep=comma] {plotdata/model_rankk_1level_naive_1core.csv};
\addplot table[x=dim,y=342,col sep=comma] {plotdata/model_rankk_1level_naive_1core.csv};
\addplot table[x=dim,y=343,col sep=comma] {plotdata/model_rankk_1level_naive_1core.csv};
\addplot table[x=dim,y=353,col sep=comma] {plotdata/model_rankk_1level_naive_1core.csv};
\addplot table[x=dim,y=363,col sep=comma] {plotdata/model_rankk_1level_naive_1core.csv};
\addplot table[x=dim,y=422,col sep=comma] {plotdata/model_rankk_1level_naive_1core.csv};
\addplot table[x=dim,y=423,col sep=comma] {plotdata/model_rankk_1level_naive_1core.csv};
\addplot table[x=dim,y=424,col sep=comma] {plotdata/model_rankk_1level_naive_1core.csv};
\addplot table[x=dim,y=432,col sep=comma] {plotdata/model_rankk_1level_naive_1core.csv};
\addplot table[x=dim,y=433,col sep=comma] {plotdata/model_rankk_1level_naive_1core.csv};
\addplot table[x=dim,y=442,col sep=comma] {plotdata/model_rankk_1level_naive_1core.csv};
\addplot table[x=dim,y=522,col sep=comma] {plotdata/model_rankk_1level_naive_1core.csv};
\addplot table[x=dim,y=633,col sep=comma] {plotdata/model_rankk_1level_naive_1core.csv};
\addplot table[x=dim,y=gemm,col sep=comma] {plotdata/model_rankk_gemm_1core.csv};
\end{axis}
\end{tikzpicture}
\caption{Performance of generated one-level ABC, AB, Naive FMM implementations on single core when \RANKK{}.
Top row: actual performance; Bottom row: modeled performance.
Left column: one-level, ABC; Middle column: one-level, AB; Right column: one-level, Naive.
}
\label{fig:rankk_1level_1core}
\end{figure*}

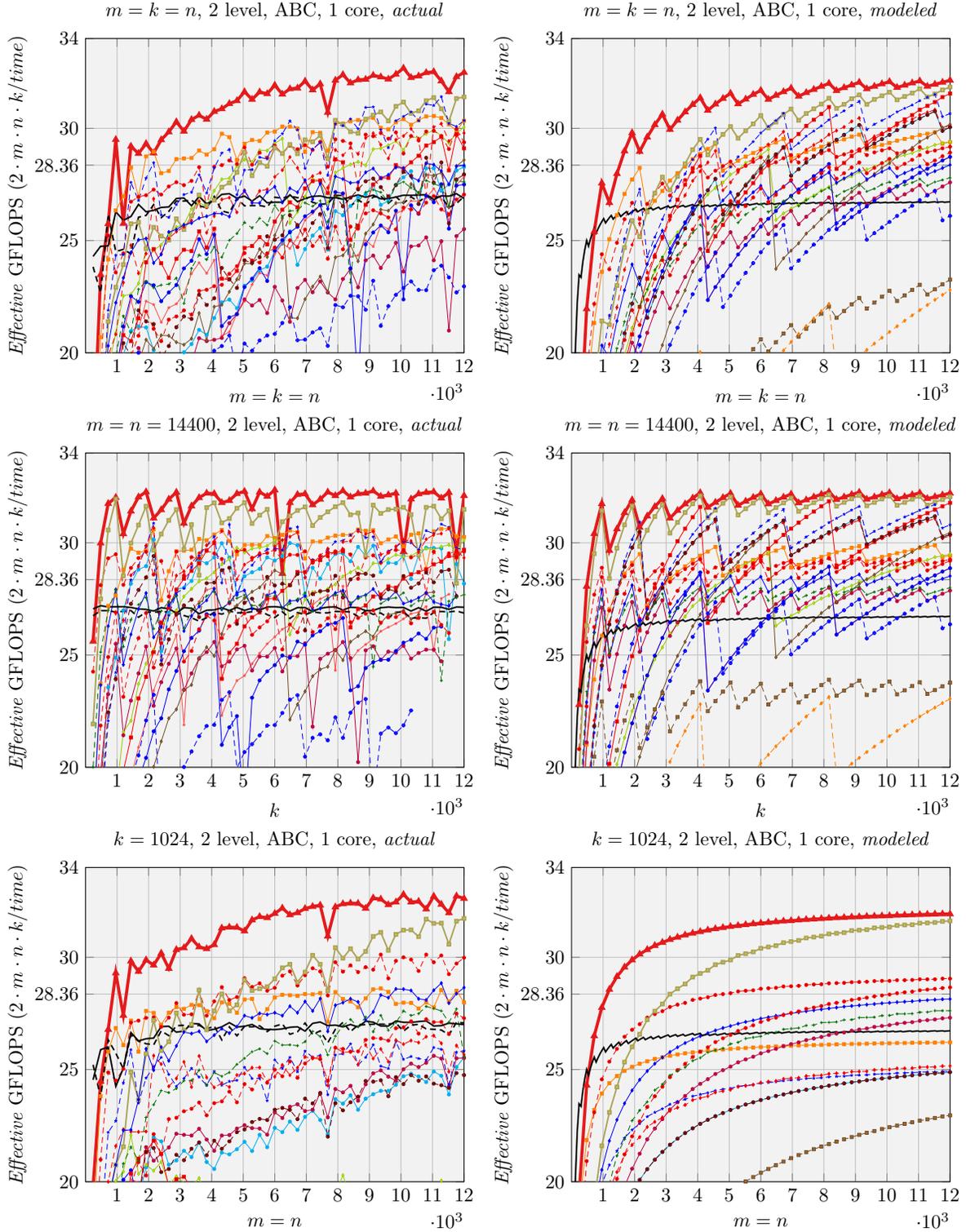
\begin{figure*}[htp!]
\center
\begin{tikzpicture}[scale=0.9]
\begin{axis}[
    title={$m = k = n$, 2 level, ABC, 1 core, \emph{actual}},
    xlabel={ $m = k = n$ },
    ylabel={\emph{Effective} GFLOPS ($ 2 \cdot  m \cdot  n \cdot k / time $)},
    xmin=0,
    xmax=12000,
    ymin=20,
    ymax=34,
    xtick={1000,2000,3000,4000,5000,6000,7000,8000,9000,10000,11000,12000},
    ytick={5, 10, 15, 20, 25, 28.36, 30, 34},
    scaled x ticks=false,
    scaled x ticks=base 10:-3,
    grid=major,
    axis background/.style={fill=lightgray!20},
    mark size=0.8pt,
    cycle list name=jianyucolor,
    restrict y to domain=1:inf,
    ]
\addplot table[x=dim,y=222,col sep=comma] {plotdata/square_2level_abc_1core.csv};
\addplot table[x=dim,y=232,col sep=comma] {plotdata/square_2level_abc_1core.csv};
\addplot table[x=dim,y=234,col sep=comma] {plotdata/square_2level_abc_1core.csv};
\addplot table[x=dim,y=243,col sep=comma] {plotdata/square_2level_abc_1core.csv};
\addplot table[x=dim,y=252,col sep=comma] {plotdata/square_2level_abc_1core.csv};
\addplot table[x=dim,y=322,col sep=comma] {plotdata/square_2level_abc_1core.csv};
\addplot table[x=dim,y=323,col sep=comma] {plotdata/square_2level_abc_1core.csv};
\addplot table[x=dim,y=324,col sep=comma] {plotdata/square_2level_abc_1core.csv};
\addplot table[x=dim,y=332,col sep=comma] {plotdata/square_2level_abc_1core.csv};
\addplot table[x=dim,y=333,col sep=comma] {plotdata/square_2level_abc_1core.csv};
\addplot table[x=dim,y=336,col sep=comma] {plotdata/square_2level_abc_1core.csv};
\addplot table[x=dim,y=342,col sep=comma] {plotdata/square_2level_abc_1core.csv};
\addplot table[x=dim,y=343,col sep=comma] {plotdata/square_2level_abc_1core.csv};
\addplot table[x=dim,y=353,col sep=comma] {plotdata/square_2level_abc_1core.csv};
\addplot table[x=dim,y=363,col sep=comma] {plotdata/square_2level_abc_1core.csv};
\addplot table[x=dim,y=422,col sep=comma] {plotdata/square_2level_abc_1core.csv};
\addplot table[x=dim,y=423,col sep=comma] {plotdata/square_2level_abc_1core.csv};
\addplot table[x=dim,y=424,col sep=comma] {plotdata/square_2level_abc_1core.csv};
\addplot table[x=dim,y=432,col sep=comma] {plotdata/square_2level_abc_1core.csv};
\addplot table[x=dim,y=433,col sep=comma] {plotdata/square_2level_abc_1core.csv};
\addplot table[x=dim,y=442,col sep=comma] {plotdata/square_2level_abc_1core.csv};
\addplot table[x=dim,y=522,col sep=comma] {plotdata/square_2level_abc_1core.csv};
\addplot table[x=dim,y=633,col sep=comma] {plotdata/square_2level_abc_1core.csv};
\addplot table[x=dim,y=blis,col sep=comma] {plotdata/square_2level_abc_1core.csv};
\addplot table[x=dim,y=mkl,col sep=comma] {plotdata/square_2level_abc_1core.csv};
\end{axis}
\end{tikzpicture}
\begin{tikzpicture}[scale=0.9]
\begin{axis}[
    title={$m = k = n$, 2 level, ABC, 1 core, \emph{modeled}},
    xlabel={ $m = k = n$ },
    ylabel={\emph{Effective} GFLOPS ($ 2 \cdot  m \cdot  n \cdot k / time $)},
    xmin=0,
    xmax=12000,
    ymin=20,
    ymax=34,
    xtick={1000,2000,3000,4000,5000,6000,7000,8000,9000,10000,11000,12000},
    ytick={5, 10, 15, 20, 25, 28.36, 30, 34},
    scaled x ticks=false,
    scaled x ticks=base 10:-3,
    grid=major,
    axis background/.style={fill=lightgray!20},
    mark size=0.8pt,
    cycle list name=jianyu0color,
    ]
\addplot table[x=dim,y=222,col sep=comma] {plotdata/model_square_2level_abc_1core.csv};
\addplot table[x=dim,y=232,col sep=comma] {plotdata/model_square_2level_abc_1core.csv};
\addplot table[x=dim,y=234,col sep=comma] {plotdata/model_square_2level_abc_1core.csv};
\addplot table[x=dim,y=243,col sep=comma] {plotdata/model_square_2level_abc_1core.csv};
\addplot table[x=dim,y=252,col sep=comma] {plotdata/model_square_2level_abc_1core.csv};
\addplot table[x=dim,y=322,col sep=comma] {plotdata/model_square_2level_abc_1core.csv};
\addplot table[x=dim,y=323,col sep=comma] {plotdata/model_square_2level_abc_1core.csv};
\addplot table[x=dim,y=324,col sep=comma] {plotdata/model_square_2level_abc_1core.csv};
\addplot table[x=dim,y=332,col sep=comma] {plotdata/model_square_2level_abc_1core.csv};
\addplot table[x=dim,y=333,col sep=comma] {plotdata/model_square_2level_abc_1core.csv};
\addplot table[x=dim,y=336,col sep=comma] {plotdata/model_square_2level_abc_1core.csv};
\addplot table[x=dim,y=342,col sep=comma] {plotdata/model_square_2level_abc_1core.csv};
\addplot table[x=dim,y=343,col sep=comma] {plotdata/model_square_2level_abc_1core.csv};
\addplot table[x=dim,y=353,col sep=comma] {plotdata/model_square_2level_abc_1core.csv};
\addplot table[x=dim,y=363,col sep=comma] {plotdata/model_square_2level_abc_1core.csv};
\addplot table[x=dim,y=422,col sep=comma] {plotdata/model_square_2level_abc_1core.csv};
\addplot table[x=dim,y=423,col sep=comma] {plotdata/model_square_2level_abc_1core.csv};
\addplot table[x=dim,y=424,col sep=comma] {plotdata/model_square_2level_abc_1core.csv};
\addplot table[x=dim,y=432,col sep=comma] {plotdata/model_square_2level_abc_1core.csv};
\addplot table[x=dim,y=433,col sep=comma] {plotdata/model_square_2level_abc_1core.csv};
\addplot table[x=dim,y=442,col sep=comma] {plotdata/model_square_2level_abc_1core.csv};
\addplot table[x=dim,y=522,col sep=comma] {plotdata/model_square_2level_abc_1core.csv};
\addplot table[x=dim,y=633,col sep=comma] {plotdata/model_square_2level_abc_1core.csv};
\addplot table[x=dim,y=gemm,col sep=comma] {plotdata/model_square_gemm_1core.csv};
\end{axis}
\end{tikzpicture}
\begin{tikzpicture}[scale=0.9]
\begin{axis}[
    title={$m = n = 14400$, 2 level, ABC, 1 core, \emph{actual}},
    xlabel={ $k$ },
    ylabel={\emph{Effective} GFLOPS ($ 2 \cdot  m \cdot  n \cdot k / time $)},
    xmin=0,
    xmax=12000,
    ymin=20,
    ymax=34,
    xtick={1000,2000,3000,4000,5000,6000,7000,8000,9000,10000,11000,12000},
    ytick={5, 10, 15, 20, 25, 28.36, 30, 34},
    scaled x ticks=false,
    scaled x ticks=base 10:-3,
    grid=major,
    axis background/.style={fill=lightgray!20},
    mark size=0.8pt,
    cycle list name=jianyucolor,
    restrict y to domain=1:inf,
    ]
\addplot table[x=dim,y=222,col sep=comma] {plotdata/rankk_2level_abc_1core.csv};
\addplot table[x=dim,y=232,col sep=comma] {plotdata/rankk_2level_abc_1core.csv};
\addplot table[x=dim,y=234,col sep=comma] {plotdata/rankk_2level_abc_1core.csv};
\addplot table[x=dim,y=243,col sep=comma] {plotdata/rankk_2level_abc_1core.csv};
\addplot table[x=dim,y=252,col sep=comma] {plotdata/rankk_2level_abc_1core.csv};
\addplot table[x=dim,y=322,col sep=comma] {plotdata/rankk_2level_abc_1core.csv};
\addplot table[x=dim,y=323,col sep=comma] {plotdata/rankk_2level_abc_1core.csv};
\addplot table[x=dim,y=324,col sep=comma] {plotdata/rankk_2level_abc_1core.csv};
\addplot table[x=dim,y=332,col sep=comma] {plotdata/rankk_2level_abc_1core.csv};
\addplot table[x=dim,y=333,col sep=comma] {plotdata/rankk_2level_abc_1core.csv};
\addplot table[x=dim,y=336,col sep=comma] {plotdata/rankk_2level_abc_1core.csv};
\addplot table[x=dim,y=342,col sep=comma] {plotdata/rankk_2level_abc_1core.csv};
\addplot table[x=dim,y=343,col sep=comma] {plotdata/rankk_2level_abc_1core.csv};
\addplot table[x=dim,y=353,col sep=comma] {plotdata/rankk_2level_abc_1core.csv};
\addplot table[x=dim,y=363,col sep=comma] {plotdata/rankk_2level_abc_1core.csv};
\addplot table[x=dim,y=422,col sep=comma] {plotdata/rankk_2level_abc_1core.csv};
\addplot table[x=dim,y=423,col sep=comma] {plotdata/rankk_2level_abc_1core.csv};
\addplot table[x=dim,y=424,col sep=comma] {plotdata/rankk_2level_abc_1core.csv};
\addplot table[x=dim,y=432,col sep=comma] {plotdata/rankk_2level_abc_1core.csv};
\addplot table[x=dim,y=433,col sep=comma] {plotdata/rankk_2level_abc_1core.csv};
\addplot table[x=dim,y=442,col sep=comma] {plotdata/rankk_2level_abc_1core.csv};
\addplot table[x=dim,y=522,col sep=comma] {plotdata/rankk_2level_abc_1core.csv};
\addplot table[x=dim,y=633,col sep=comma] {plotdata/rankk_2level_abc_1core.csv};
\addplot table[x=dim,y=blis,col sep=comma] {plotdata/rankk_2level_abc_1core.csv};
\addplot table[x=dim,y=mkl,col sep=comma] {plotdata/rankk_2level_abc_1core.csv};
\end{axis}
\end{tikzpicture}
\begin{tikzpicture}[scale=0.9]
\begin{axis}[
    title={$m = n = 14400$, 2 level, ABC, 1 core, \emph{modeled}},
    xlabel={ $k$ },
    ylabel={\emph{Effective} GFLOPS ($ 2 \cdot  m \cdot  n \cdot k / time $)},
    xmin=0,
    xmax=12000,
    ymin=20,
    ymax=34,
    xtick={1000,2000,3000,4000,5000,6000,7000,8000,9000,10000,11000,12000},
    ytick={5, 10, 15, 20, 25, 28.36, 30, 34},
    scaled x ticks=false,
    scaled x ticks=base 10:-3,
    grid=major,
    axis background/.style={fill=lightgray!20},
    mark size=0.8pt,
    cycle list name=jianyu0color,
    ]
\addplot table[x=dim,y=222,col sep=comma] {plotdata/model_rankk_2level_abc_1core.csv};
\addplot table[x=dim,y=232,col sep=comma] {plotdata/model_rankk_2level_abc_1core.csv};
\addplot table[x=dim,y=234,col sep=comma] {plotdata/model_rankk_2level_abc_1core.csv};
\addplot table[x=dim,y=243,col sep=comma] {plotdata/model_rankk_2level_abc_1core.csv};
\addplot table[x=dim,y=252,col sep=comma] {plotdata/model_rankk_2level_abc_1core.csv};
\addplot table[x=dim,y=322,col sep=comma] {plotdata/model_rankk_2level_abc_1core.csv};
\addplot table[x=dim,y=323,col sep=comma] {plotdata/model_rankk_2level_abc_1core.csv};
\addplot table[x=dim,y=324,col sep=comma] {plotdata/model_rankk_2level_abc_1core.csv};
\addplot table[x=dim,y=332,col sep=comma] {plotdata/model_rankk_2level_abc_1core.csv};
\addplot table[x=dim,y=333,col sep=comma] {plotdata/model_rankk_2level_abc_1core.csv};
\addplot table[x=dim,y=336,col sep=comma] {plotdata/model_rankk_2level_abc_1core.csv};
\addplot table[x=dim,y=342,col sep=comma] {plotdata/model_rankk_2level_abc_1core.csv};
\addplot table[x=dim,y=343,col sep=comma] {plotdata/model_rankk_2level_abc_1core.csv};
\addplot table[x=dim,y=353,col sep=comma] {plotdata/model_rankk_2level_abc_1core.csv};
\addplot table[x=dim,y=363,col sep=comma] {plotdata/model_rankk_2level_abc_1core.csv};
\addplot table[x=dim,y=422,col sep=comma] {plotdata/model_rankk_2level_abc_1core.csv};
\addplot table[x=dim,y=423,col sep=comma] {plotdata/model_rankk_2level_abc_1core.csv};
\addplot table[x=dim,y=424,col sep=comma] {plotdata/model_rankk_2level_abc_1core.csv};
\addplot table[x=dim,y=432,col sep=comma] {plotdata/model_rankk_2level_abc_1core.csv};
\addplot table[x=dim,y=433,col sep=comma] {plotdata/model_rankk_2level_abc_1core.csv};
\addplot table[x=dim,y=442,col sep=comma] {plotdata/model_rankk_2level_abc_1core.csv};
\addplot table[x=dim,y=522,col sep=comma] {plotdata/model_rankk_2level_abc_1core.csv};
\addplot table[x=dim,y=633,col sep=comma] {plotdata/model_rankk_2level_abc_1core.csv};
\addplot table[x=dim,y=gemm,col sep=comma] {plotdata/model_rankk_gemm_1core.csv};
\end{axis}
\end{tikzpicture}
\begin{tikzpicture}[scale=0.9]
\begin{axis}[
    title={$k = 1024$, 2 level, ABC, 1 core, \emph{actual}},
    xlabel={ $m = n$ },
    ylabel={\emph{Effective} GFLOPS ($ 2 \cdot  m \cdot  n \cdot k / time $)},
    xmin=0,
    xmax=12000,
    ymin=20,
    ymax=34,
    xtick={1000,2000,3000,4000,5000,6000,7000,8000,9000,10000,11000,12000},
    ytick={5, 10, 15, 20, 25, 28.36, 30, 34},
    scaled x ticks=false,
    scaled x ticks=base 10:-3,
    grid=major,
    axis background/.style={fill=lightgray!20},
    mark size=0.8pt,
    cycle list name=jianyucolor,
    restrict y to domain=1:inf,
    ]
\addplot table[x=dim,y=222,col sep=comma] {plotdata/fixk_2level_abc_1core.csv};
\addplot table[x=dim,y=232,col sep=comma] {plotdata/fixk_2level_abc_1core.csv};
\addplot table[x=dim,y=234,col sep=comma] {plotdata/fixk_2level_abc_1core.csv};
\addplot table[x=dim,y=243,col sep=comma] {plotdata/fixk_2level_abc_1core.csv};
\addplot table[x=dim,y=252,col sep=comma] {plotdata/fixk_2level_abc_1core.csv};
\addplot table[x=dim,y=322,col sep=comma] {plotdata/fixk_2level_abc_1core.csv};
\addplot table[x=dim,y=323,col sep=comma] {plotdata/fixk_2level_abc_1core.csv};
\addplot table[x=dim,y=324,col sep=comma] {plotdata/fixk_2level_abc_1core.csv};
\addplot table[x=dim,y=332,col sep=comma] {plotdata/fixk_2level_abc_1core.csv};
\addplot table[x=dim,y=333,col sep=comma] {plotdata/fixk_2level_abc_1core.csv};
\addplot table[x=dim,y=336,col sep=comma] {plotdata/fixk_2level_abc_1core.csv};
\addplot table[x=dim,y=342,col sep=comma] {plotdata/fixk_2level_abc_1core.csv};
\addplot table[x=dim,y=343,col sep=comma] {plotdata/fixk_2level_abc_1core.csv};
\addplot table[x=dim,y=353,col sep=comma] {plotdata/fixk_2level_abc_1core.csv};
\addplot table[x=dim,y=363,col sep=comma] {plotdata/fixk_2level_abc_1core.csv};
\addplot table[x=dim,y=422,col sep=comma] {plotdata/fixk_2level_abc_1core.csv};
\addplot table[x=dim,y=423,col sep=comma] {plotdata/fixk_2level_abc_1core.csv};
\addplot table[x=dim,y=424,col sep=comma] {plotdata/fixk_2level_abc_1core.csv};
\addplot table[x=dim,y=432,col sep=comma] {plotdata/fixk_2level_abc_1core.csv};
\addplot table[x=dim,y=433,col sep=comma] {plotdata/fixk_2level_abc_1core.csv};
\addplot table[x=dim,y=442,col sep=comma] {plotdata/fixk_2level_abc_1core.csv};
\addplot table[x=dim,y=522,col sep=comma] {plotdata/fixk_2level_abc_1core.csv};
\addplot table[x=dim,y=633,col sep=comma] {plotdata/fixk_2level_abc_1core.csv};
\addplot table[x=dim,y=blis,col sep=comma] {plotdata/fixk_2level_abc_1core.csv};
\addplot table[x=dim,y=mkl,col sep=comma] {plotdata/fixk_2level_abc_1core.csv};
\end{axis}
\end{tikzpicture}
\begin{tikzpicture}[scale=0.9]
\begin{axis}[
    title={$k = 1024$, 2 level, ABC, 1 core, \emph{modeled}},
    xlabel={ $m = n$ },
    ylabel={\emph{Effective} GFLOPS ($ 2 \cdot  m \cdot  n \cdot k / time $)},
    xmin=0,
    xmax=12000,
    ymin=20,
    ymax=34,
    xtick={1000,2000,3000,4000,5000,6000,7000,8000,9000,10000,11000,12000},
    ytick={5, 10, 15, 20, 25, 28.36, 30, 34},
    scaled x ticks=false,
    scaled x ticks=base 10:-3,
    grid=major,
    axis background/.style={fill=lightgray!20},
    mark size=0.8pt,
    cycle list name=jianyu0color,
    ]
\addplot table[x=dim,y=222,col sep=comma] {plotdata/model_fixk_2level_abc_1core.csv};
\addplot table[x=dim,y=232,col sep=comma] {plotdata/model_fixk_2level_abc_1core.csv};
\addplot table[x=dim,y=234,col sep=comma] {plotdata/model_fixk_2level_abc_1core.csv};
\addplot table[x=dim,y=243,col sep=comma] {plotdata/model_fixk_2level_abc_1core.csv};
\addplot table[x=dim,y=252,col sep=comma] {plotdata/model_fixk_2level_abc_1core.csv};
\addplot table[x=dim,y=322,col sep=comma] {plotdata/model_fixk_2level_abc_1core.csv};
\addplot table[x=dim,y=323,col sep=comma] {plotdata/model_fixk_2level_abc_1core.csv};
\addplot table[x=dim,y=324,col sep=comma] {plotdata/model_fixk_2level_abc_1core.csv};
\addplot table[x=dim,y=332,col sep=comma] {plotdata/model_fixk_2level_abc_1core.csv};
\addplot table[x=dim,y=333,col sep=comma] {plotdata/model_fixk_2level_abc_1core.csv};
\addplot table[x=dim,y=336,col sep=comma] {plotdata/model_fixk_2level_abc_1core.csv};
\addplot table[x=dim,y=342,col sep=comma] {plotdata/model_fixk_2level_abc_1core.csv};
\addplot table[x=dim,y=343,col sep=comma] {plotdata/model_fixk_2level_abc_1core.csv};
\addplot table[x=dim,y=353,col sep=comma] {plotdata/model_fixk_2level_abc_1core.csv};
\addplot table[x=dim,y=363,col sep=comma] {plotdata/model_fixk_2level_abc_1core.csv};
\addplot table[x=dim,y=422,col sep=comma] {plotdata/model_fixk_2level_abc_1core.csv};
\addplot table[x=dim,y=423,col sep=comma] {plotdata/model_fixk_2level_abc_1core.csv};
\addplot table[x=dim,y=424,col sep=comma] {plotdata/model_fixk_2level_abc_1core.csv};
\addplot table[x=dim,y=432,col sep=comma] {plotdata/model_fixk_2level_abc_1core.csv};
\addplot table[x=dim,y=433,col sep=comma] {plotdata/model_fixk_2level_abc_1core.csv};
\addplot table[x=dim,y=442,col sep=comma] {plotdata/model_fixk_2level_abc_1core.csv};
\addplot table[x=dim,y=522,col sep=comma] {plotdata/model_fixk_2level_abc_1core.csv};
\addplot table[x=dim,y=633,col sep=comma] {plotdata/model_fixk_2level_abc_1core.csv};
\addplot table[x=dim,y=gemm,col sep=comma] {plotdata/model_fixk_gemm_1core.csv};
\end{axis}
\end{tikzpicture}
\caption{Performance of generated two-level ABC FMM implementations on single core when \SQUARE{}; \RANKK{}; \FIXK{}.
Top row: actual performance; Bottom row: modeled performance.
Left column: \SQUARE{}; Middle column: \RANKK{}; Right column: \FIXK{}.
}
\label{fig:2level_abc_1core}
\end{figure*}

In \cite{Strassen:SC16}, a performance model was given to estimate the execution time $ T $
for the one-level/two-level
ABC, AB, and Naive variations of $ \langle 2, 2, 2 \rangle $ \strassen{}.
In this subsection, we  generalize that performance model to predict the execution time $ T $
for the various FMM implementations generated by our code generator.
Theoretical estimation helps us better understand the computation and memory footprint of different FMM implementations,
and allows us to avoid exhaustive empirical search when searching for the best implementation for different problem sizes and shapes.
Most importantly, our code generator can embed our performance model to guide the selection of a FMM implementation as a function of problem size and shape,
with the input $ \langle \widetilde m_l, \widetilde k_l, \widetilde n_l \rangle $ and $ \llbracket U_l, V_l, W_l \rrbracket $ specifications on each level $l$.
These performance models are themselves automatically generated.


\subsubsection*{Assumption}
We have similar architecture assumptions as in \cite{Strassen:SC16}.
Basically we assume that the architecture has two layers of modern memory hierarchy: fast caches and relatively slow main memory (DRAM).
For read operations, the latency for accessing cache can be ignored, while the latency for accessing the main memory is counted;
For write operations, we assume a lazy write-back policy such that the time for writing into fast caches can be hidden.
Based on these  assumptions, the memory operations for \gemm{} and various implementations of FMM are decomposed into three parts:
\begin{itemize}
\item memory packing shown in \figref{fig:side_by_side}.
\item reading/writing the submatrices of $ C $ in \figref{fig:side_by_side}.
\item reading/writing of the temporary buffer that are parts of \XXXFMM{}/\ABXFMM{}.
\end{itemize}

\subsubsection*{Notation}
Notation is summarized in \figref{tab:notation}.

The total execution time, $ T $, is dominated by arithmetic time $ T_a $ and
memory time $ T_m $ (\mycircle{2} in \figref{tab:perfmodel}).

\subsubsection*{Arithmetic operations}

$ T_a $ is decomposed into submatrix multiplications ($T_{a}^{\times}$) and submatrix additions ($T_{a}^{A_{+}}$, $T_{a}^{B_{+}}$, $T_{a}^{C_{+}}$) (\mycircle{3} in \figref{tab:perfmodel}).
$ T_{a}^{X_{+}} $ has a coefficient 2 because under the hood the matrix additions are cast into {\tt FMA} operations.
The corresponding coefficients $ N_a^X $ are tabulated in \figref{tab:perfmodel}. For instance,
$ N_{a}^{A_{+}}$ = $nnz(\OTIMESU{})-\PRODRL{} $ for $L$-level FMM, because
computing $ \sum { ((\OTIMESU{})_{i,r} {A}_{i}) } $ in (\ref{eqn:multifmm}) involves
$ \sum\nolimits_{r=0}^{R_L-1}(nnz((\OTIMESU{})_{:,r}) - 1)$ = $nnz(\OTIMESU{})-\!\PRODRL{} $ submatrix additions.
Note that $ X_{:,r} $ denotes the $ r $th column of $ X $.

\subsubsection*{Memory operations}
$ T_m $ is a function of
the submatrix sizes \{$ m/\widetilde{M_L} $, $ k/\widetilde{K_L} $, $ n/\widetilde{N_L} $\},
and the block sizes \{$ m_C $, $ k_C $, $n_C$\} in \figref{fig:side_by_side}(right), because the memory operation can repeat multiple times according to which loop they reside in.
$ T_m $ is broken down into several components, as shown in \mycircle{4} in \figref{tab:perfmodel}.
Each memory operation term is characterized in \figref{tab:perfmodel} by its
read/write type and the amount of memory in units of 64-bit double precision elements.
Note that
$T_{m}^{{\widetilde A}_{\times}}$,$T_{m}^{{\widetilde B}_{\times}}$ are omitted in \mycircle{4} because of the assumption of lazy write-back policy with fast caches.
Due to the software prefetching effects,
$T_{m}^{C_{\times}} \!\!=\!\! 2\lambda \frac{m}{\DivisorML{}}\frac{n}{\DivisorNL{}}\lceil\frac{k/\DivisorKL{}}{k_c}\rceil \tau_b $
has an additional parameter $\lambda\in[0.5,1]$, which denotes the prefetching efficiency. $\lambda$ is adapted to match \gemm{} performance. Note that this is a ceiling function proportional to $ k $, because rank-k updates for accumulating submatrices of $ C $ recur $ \lceil\frac{k/\DivisorKL{}}{k_c}\rceil $ times in 4th loop in \figref{fig:side_by_side}.
The corresponding coefficients $ N_m^X $ are tabulated in \figref{tab:perfmodel}.
For example, for \XXXFMM{} and \ABXFMM{}, computing
$ {C}_{p} +\!\!= (\OTIMESW{})_{p,r} {M}_{r} ( p = 0,... ) $ in (\ref{eqn:multifmm})
involves 2 read and 1 write related to temporary buffer in slow memory.
Therefore,
$ N_{m}^{C_{\times}}$ = $3nnz(\OTIMESW{}) $.



\subsection{Discussion} \label{s:model_discussion}
We can make estimations about the run time performance of the various FMM implementations generated by our code generator,
based on the analysis shown in \figref{tab:perfmodel}.
We use \emph{Effective} GFLOPS (defined in \mycircle{1} in \figref{tab:perfmodel}) as the metric to compare the performance of these various FMM implementations, similar to \cite{Benson15,PPL2,StrassenLipshitz}.
The architecture-dependent parameters for the model are given in Section \ref{s:single_node}.
We demonstrate the performance of two representative groups of experiments in Figures \ref{fig:rankk_1level_1core} and \ref{fig:2level_abc_1core}.

\begin{itemize}


%
%


\item
Contrary to what was observed in \cite{Strassen:SC16}, \XXXFMM{} may perform better than \ABCFMM{} and \ABXFMM{} for relatively large problem size.
For example, in \figref{fig:rankk_1level_1core}, $ \langle 3, 6, 3 \rangle $ (with the maximum theoretical speedup among all FMMs we test, \figref{tab:fmm_alg}) has better \XXXFMM{} performance than \ABCFMM{} and \ABXFMM{}.
This is because the total number of times for packing in $ \langle 3, 6, 3 \rangle $ is very large
($N_{m}^{{A}_{\times}} = nnz(\OTIMESU{}) $, $N_{m}^{{B}_{\times}} = nnz(\OTIMESV{}) $).
This magnifies the overhead for packing with \ABXFMM{}/\ABCFMM{}.

\item
Contrary to what was observed in \cite{Benson15},
for rank-k updates (middle column, right column, \figref{fig:2level_abc_1core}), $ \langle 2, 2, 2 \rangle $ still performs the best with \ABCFMM{} implementations (\cite{Benson15} observe some other shapes, e.g. $ \langle 4, 2, 4 \rangle $, tend to have higher performance). This is because their implementations are similar to \XXXFMM{}, with the overhead for forming the $ M_r $ matrices explicitly.

\item
\figref{fig:rankk_1level_1core} shows that
for small problem size,
when $ k $ is small, \ABCFMM{} performs best;
when $ k $ is large, \ABXFMM{}/\XXXFMM{} perform better.
That can be quantitatively explained by comparing the coefficients of $N_m^X$ in the bottom table in \figref{tab:perfmodel}.

\item
The graph for \mbox{$ m = n = 14400$, $k$ varies, ABC, 1core} (left column, \figref{fig:rankk_1level_1core}; middle column, \figref{fig:2level_abc_1core})
shows that 
for $ k $ equal to the appropriate multiple of $ k_C $ (
$ k =  \prod\nolimits_{l=0}^{L-1}{\widetilde {k_l}}  \times k_C $), \ABCFMM{}
achieves the best performance.

\end{itemize}


\subsection{Apply performance model to code generator}


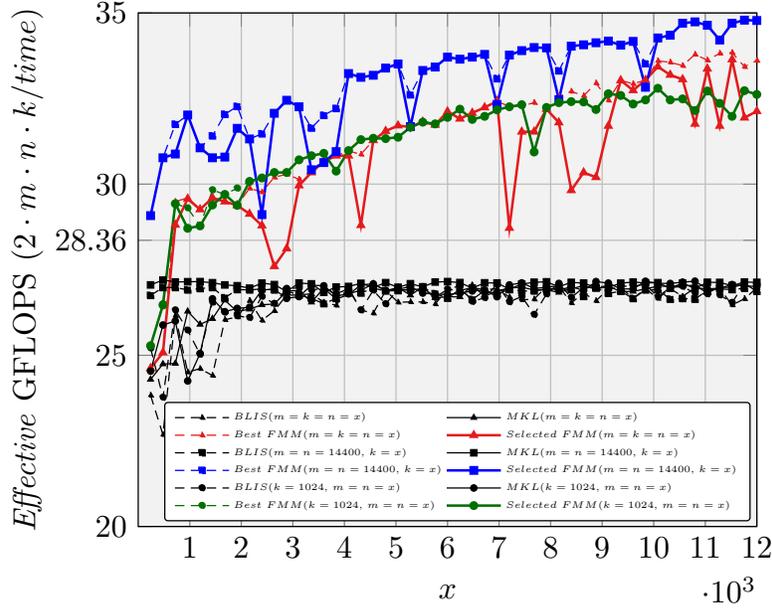
\begin{figure}[tp!]
\center 
\begin{tikzpicture}[scale=1.2]
\begin{axis}[
    title={},
    xlabel={ $ x $ },
    ylabel={\emph{Effective} GFLOPS ($ 2 \cdot  m \cdot  n \cdot k / time $)},
    xmin=0,
    xmax=12000,
    ymin=20,
    ymax=35,
    xtick={1000,2000,3000,4000,5000,6000,7000,8000,9000,10000,11000,12000},
    ytick={5, 10, 15, 20, 25, 28.36, 30, 35},
    scaled x ticks=false,
    scaled x ticks=base 10:-3,
    grid=major,
    axis background/.style={fill=lightgray!20},
    mark size=1pt,
    cycle list name=jianyu4color,
    restrict y to domain=1:inf,
    legend style={
        at={(0.99,0.01)},
        anchor=south east,
        legend columns=2,
        font=\tiny,
        rounded corners=1pt,
        nodes={scale=0.6, transform shape},
        cells={anchor=west},
    },
    legend entries = {
        $BLIS$($m=k=n=x$)\\
        $MKL$($m=k=n=x$)\\
        $Best~FMM$($m=k=n=x$)\\
        $Selected~FMM$($m=k=n=x$)\\
        $BLIS$($m=n=14400$, $k=x$)\\
        $MKL$($m=n=14400$, $k=x$)\\
        $Best~FMM$($m=n=14400$, $k=x$)\\
        $Selected~FMM$($m=n=14400$, $k=x$)\\
        $BLIS$($k=1024$, $m=n=x$)\\
        $MKL$($k=1024$, $m=n=x$)\\
        $Best~FMM$($k=1024$, $m=n=x$)\\
        $Selected~FMM$($k=1024$, $m=n=x$)\\
    },
    ]
\addplot table[x=dim,y=blis,col sep=comma] {plotdata/maxline2_square_1core.csv};
\addplot table[x=dim,y=mkl,col sep=comma] {plotdata/maxline2_square_1core.csv};
\addplot table[x=dim,y=maxline,col sep=comma] {plotdata/maxline2_square_1core.csv};
\addplot table[x=dim,y=predicted_maxline,col sep=comma] {plotdata/maxline2_square_1core.csv};

\addplot table[x=dim,y=blis,col sep=comma] {plotdata/maxline2_rankk_1core.csv};
\addplot table[x=dim,y=mkl,col sep=comma] {plotdata/maxline2_rankk_1core.csv};
\addplot table[x=dim,y=maxline,col sep=comma] {plotdata/maxline2_rankk_1core.csv};
\addplot table[x=dim,y=predicted_maxline,col sep=comma] {plotdata/maxline2_rankk_1core.csv};

\addplot table[x=dim,y=blis,col sep=comma] {plotdata/maxline2_fixk_1core.csv};
\addplot table[x=dim,y=mkl,col sep=comma] {plotdata/maxline2_fixk_1core.csv};
\addplot table[x=dim,y=maxline,col sep=comma] {plotdata/maxline2_fixk_1core.csv};
\addplot table[x=dim,y=predicted_maxline,col sep=comma] {plotdata/maxline2_fixk_1core.csv};

\end{axis}
\end{tikzpicture}
\caption{Selecting FMM implementation with performance model.}
\label{fig:apply_model}
\end{figure}


For actual performance, even the best implementation has some unexpected drops,
due to the ``fringes'' which are caused by the problem sizes not being divisible by
partition dimesions $\widetilde m$, $\widetilde k$, $\widetilde n$.
This is not captured by our performance model.
Therefore, given the specific problem size and shape, we choose the best two implementations
predicted by our performance model as the top two candidate implementations,
and then measure the performance in practice to pick the best one.

In \figref{fig:apply_model} we show the performance results on single core by selecting the generated FMM implementation with the guide of performance model,
when \SQUARE{}; \RANKK{}; and \FIXK{}.

Overall this experiment shows that the performance model is accurate enough in terms of
relative performance between various FMM implementations to guide the choice of a FMM implementation,
with the problem sizes and shapes as the inputs.
That will reduce the potential overhead of exhaustive empirical search.

\NoShow{
\subsection{Parallelization}

It is trivial to parallelize Xeon CPU like Ivy bridge.

Discussion of parallel scheme for diff levels

KNL parallelizes $ j_C $ and $ i_C $ loop, which is non-trivial.

}

\section{Performance Experiments} \label{s:perf}

We present performance evaluations for various generated FMM implementations.
\NoShow{The code generator can generate code for Intel Sandy-Bridge/Ivy-Bridge processor and
second-generation Intel Xeon Phi coprocessor (KNL).}

\label{s:single_node}

\subsection{Implementation and architecture information} 
The FMM implementations generated by our code generator are written in {\tt C}, utilizing {\tt SSE2} and {\tt AVX} assembly, 
compiled with the Intel {\tt C} compiler version 15.0.3 with optimization flag {\tt -O3 -mavx}. 

We compare against our generated \dgemm{} (based on the packing routines and micro-kernel borrowed from \BLIS{}, marked as BLIS in the performance figures) as well as 
Intel MKL's \dgemm{} \cite{IntelMKL} (marked as MKL in the performance figures).


We measure performance on a  
dual-socket (10 cores/socket) Intel Xeon E5-2680 v2 (Ivy Bridge) processor with 12.8 GB/core of memory
(Peak Bandwidth: 59.7 GB/s with four channels) and a three-level cache: 32 KB L1 data cache, 256 KB L2 cache and 25.6 MB L3 cache.
The stable CPU clockrate is 3.54 GHz when a single core is utilized (28.32 GFLOPS peak, marked in the graphs) and 3.10 GHz when ten cores are in use (24.8 GLOPS/core peak). 
To set thread affinity and to ensure the computation and the memory allocation all reside on 
the same socket, we use {\tt KMP\_AFFINITY=compact}.

The blocking parameters, $n_R=4$, $m_R=8$, $k_C=256$, $n_C=4096$ and $m_C=96$,
are consistent with parameters used for the standard \BLIS\ \dgemm\ implementation for this architecture.
This makes the size of the packing buffer
$ \widetilde A_i $ 192 KB and 
$ \widetilde B_p $ 8192 KB, which then fit the L2 cache and L3 cache, respectively.

\NoShow{Each socket consists of 10 cores, allowing us to also perform multi-threaded experiments.}
Parallelization 
is implemented mirroring that described in~\cite{BLIS3}, using OpenMP directives that parallelize the third loop around the micro-kernel in Figure~\ref{fig:side_by_side}.  

\FromTo{\subsubsection*{Benefit of hybrid partitions}}{\subsection{Benefit of hybrid partitions}}
First, we demonstrate the benefit of using different FMM algorithms for each level.

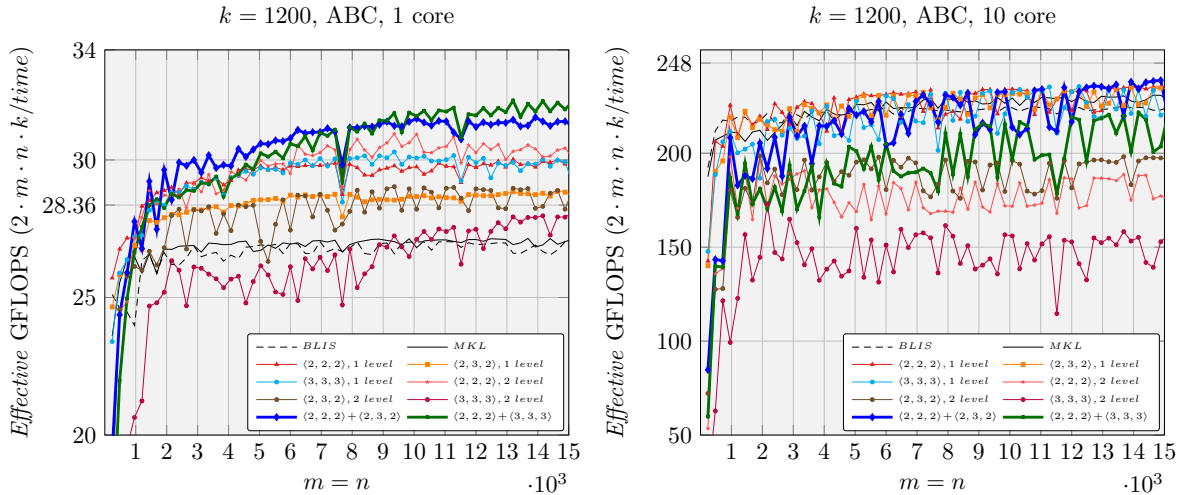
\begin{figure*}[t!]
\center 
\begin{tikzpicture}[scale=0.9]
\begin{axis}[
    title={$k = 1200$, ABC, 1 core},
    xlabel={ $m = n$ },
    ylabel={\emph{Effective} GFLOPS ($ 2 \cdot  m \cdot  n \cdot k / time $)},
    xmin=0,
    xmax=15000,
    ymin=20,
    ymax=34,
    xtick={1000,2000,3000,4000,5000,6000,7000,8000,9000,10000,11000,12000,13000,14000,15000},
    ytick={5, 10, 15, 20, 25, 28.36, 30, 34},
    scaled x ticks=false,
    scaled x ticks=base 10:-3,
    grid=major,
    axis background/.style={fill=lightgray!20},
    mark size=0.8pt,
    cycle list name=jianyu3color,
    legend style={
        at={(0.99,0.01)},
        anchor=south east,
        legend columns=2,
        font=\tiny,
        rounded corners=1pt,
        nodes={scale=0.8, transform shape},
        cells={anchor=west},
    },
    legend entries = {
        $ BLIS    $\\
        $ MKL     $\\
        $ \langle 2, 2, 2 \rangle, 1~level$\\
        $ \langle 2, 3, 2 \rangle, 1~level$\\
        $ \langle 3, 3, 3 \rangle, 1~level$\\
        $ \langle 2, 2, 2 \rangle, 2~level$\\ 
        $ \langle 2, 3, 2 \rangle, 2~level$\\ 
        $ \langle 3, 3, 3 \rangle, 2~level$\\
        $ \langle 2, 2, 2 \rangle \!+\! \langle 2, 3, 2 \rangle $\\ 
        $ \langle 2, 2, 2 \rangle \!+\! \langle 3, 3, 3 \rangle $\\ 
    },
    ]
\addplot table[x=dim,y=blis,col sep=comma] {plotdata/fixk1200_1core.csv};
\addplot table[x=dim,y=mkl,col sep=comma] {plotdata/fixk1200_1core.csv};
\addplot table[x=dim,y=222_1,col sep=comma] {plotdata/fixk1200_1core.csv};
\addplot table[x=dim,y=232_1,col sep=comma] {plotdata/fixk1200_1core.csv};
\addplot table[x=dim,y=333_1,col sep=comma] {plotdata/fixk1200_1core.csv};
\addplot table[x=dim,y=222_2,col sep=comma] {plotdata/fixk1200_1core.csv};
\addplot table[x=dim,y=232_2,col sep=comma] {plotdata/fixk1200_1core.csv};
\addplot table[x=dim,y=333_2,col sep=comma] {plotdata/fixk1200_1core.csv};
\addplot table[x=dim,y=222+232,col sep=comma] {plotdata/fixk1200_1core.csv};
\addplot table[x=dim,y=222+333,col sep=comma] {plotdata/fixk1200_1core.csv};
\end{axis}
\end{tikzpicture}
\begin{tikzpicture}[scale=0.9]
\begin{axis}[
    title={$k = 1200$, ABC, 10 core},
    xlabel={ $m = n$ },
    ylabel={\emph{Effective} GFLOPS ($ 2 \cdot  m \cdot  n \cdot k / time $)},
    xmin=0,
    xmax=15000,
    ymin=50,
    ymax=255,
    xtick={1000,2000,3000,4000,5000,6000,7000,8000,9000,10000,11000,12000,13000,14000,15000,16000},
    ytick={50, 100, 150, 200, 248},
    scaled x ticks=false,
    scaled x ticks=base 10:-3,
    grid=major,
    axis background/.style={fill=lightgray!20},
    mark size=0.8pt,
    cycle list name=jianyu3color,
    legend style={
        at={(0.99,0.01)},
        anchor=south east,
        legend columns=2,
        font=\tiny,
        rounded corners=1pt,
        nodes={scale=0.8, transform shape},
        cells={anchor=west},
    },
    legend entries = {
        $ BLIS    $\\
        $ MKL     $\\
        $ \langle 2, 2, 2 \rangle, 1~level$\\
        $ \langle 2, 3, 2 \rangle, 1~level$\\
        $ \langle 3, 3, 3 \rangle, 1~level$\\
        $ \langle 2, 2, 2 \rangle, 2~level$\\ 
        $ \langle 2, 3, 2 \rangle, 2~level$\\ 
        $ \langle 3, 3, 3 \rangle, 2~level$\\
        $ \langle 2, 2, 2 \rangle \!+\! \langle 2, 3, 2 \rangle $\\ 
        $ \langle 2, 2, 2 \rangle \!+\! \langle 3, 3, 3 \rangle $\\ 
    },
    ]
\addplot table[x=dim,y=blis,col sep=comma] {plotdata/fixk1200_10core.csv};
\addplot table[x=dim,y=mkl,col sep=comma] {plotdata/fixk1200_10core.csv};
\addplot table[x=dim,y=222_1,col sep=comma] {plotdata/fixk1200_10core.csv};
\addplot table[x=dim,y=232_1,col sep=comma] {plotdata/fixk1200_10core.csv};
\addplot table[x=dim,y=333_1,col sep=comma] {plotdata/fixk1200_10core.csv};
\addplot table[x=dim,y=222_2,col sep=comma] {plotdata/fixk1200_10core.csv};
\addplot table[x=dim,y=232_2,col sep=comma] {plotdata/fixk1200_10core.csv};
\addplot table[x=dim,y=333_2,col sep=comma] {plotdata/fixk1200_10core.csv};
\addplot table[x=dim,y=222+232,col sep=comma] {plotdata/fixk1200_10core.csv};
\addplot table[x=dim,y=222+333,col sep=comma] {plotdata/fixk1200_10core.csv};
\end{axis}
\end{tikzpicture}
\caption{Benefit of hybrid partitions over other partitions.}
\label{fig:hybrid}
\end{figure*}

\NoShow{Hybrid partitions can be easily expressed by our multi-level FMM representation with Kronecker product.}

We report the performance of different combinations of one-level/two-level
$ \langle 2,2,2 \rangle $,
$ \langle 2,3,2 \rangle $, and
$ \langle 3,3,3 \rangle $
in 
\figref{fig:hybrid}, when $ k $ is fixed to 1200 and $ m = n $ vary.
As suggested and illustrated in Section \ref{s:model_discussion}, \ABCFMM{} performs best for rank-k updates, which is why we only show the \ABCFMM{} performance.

Overall the hybrid partitions
$ \langle 2,2,2 \rangle $ + $ \langle 2,3,2 \rangle $
and
$ \langle 2,2,2 \rangle $ + $ \langle 3,3,3 \rangle $
achieve the best performance.
This is because 1200 is close to $ 2 \times 3 \times k_C $, meaning that the hybrid partitions of 2 and 3 on the $ k $ dimension are more favorable.
This is consistent with what the performance model predicts.
Performance benefits are less
for 10 cores due to bandwidth limitations,
although performance of hybrid partitions still beats two-level homogeneous partitions.

This experiment shows
the benefit of hybrid partitions, 
facilitated by the Kronecker product representation.


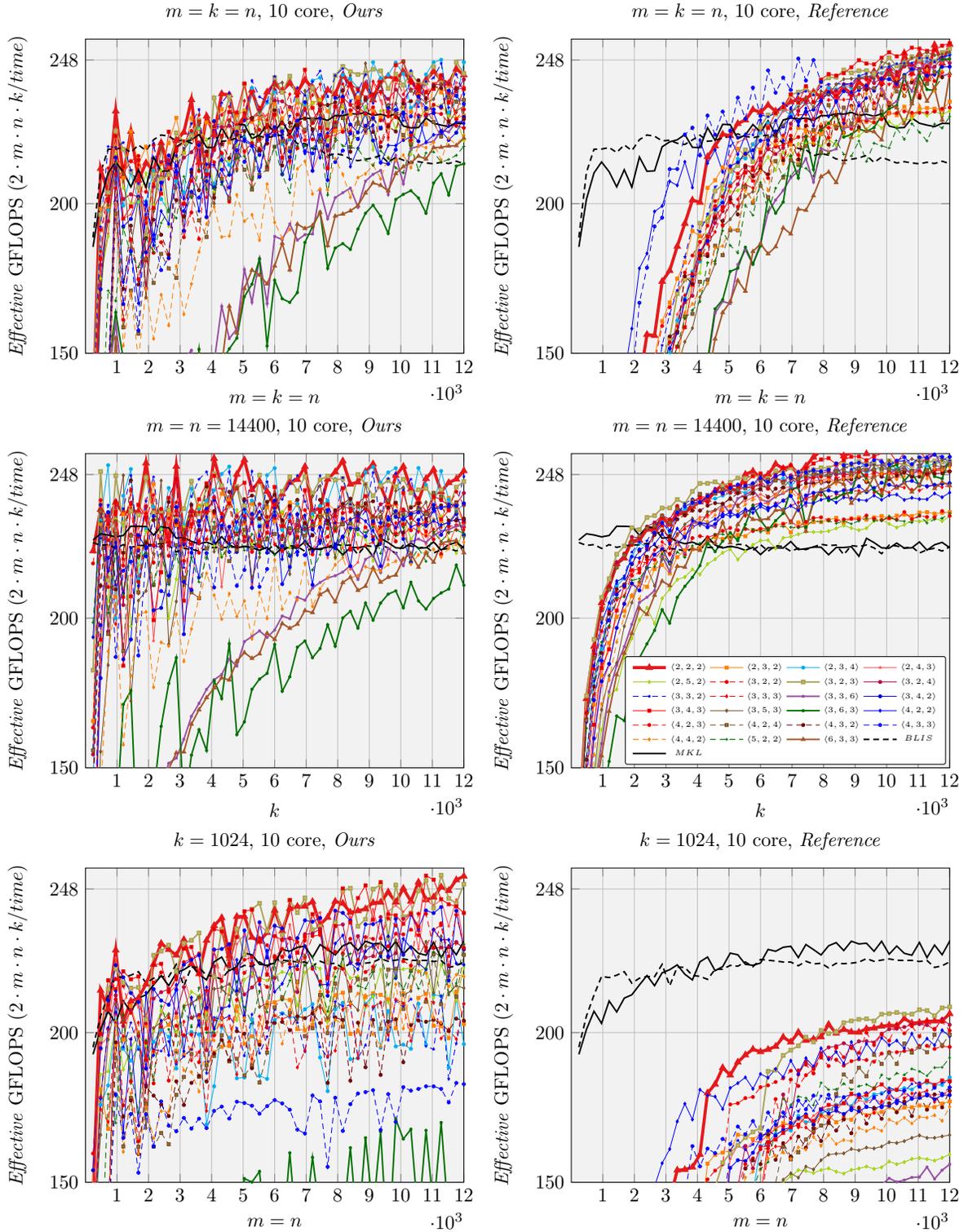
\begin{figure*}[p]
\center
\begin{tikzpicture}[scale=.9]
\begin{axis}[
    title={$m = k = n$, 10 core, \emph{Ours}},
    xlabel={ $m = k = n$ },
    ylabel={\emph{Effective} GFLOPS ($ 2 \cdot  m \cdot  n \cdot k / time $)},
    xmin=0,
    xmax=12000,
    ymin=150,
    ymax=255,
    xtick={1000,2000,3000,4000,5000,6000,7000,8000,9000,10000,11000,12000},
    ytick={50, 100, 150, 200, 248},
    scaled x ticks=false,
    scaled x ticks=base 10:-3,
    grid=major,
    axis background/.style={fill=lightgray!20},
    mark size=0.8pt,
    cycle list name=jianyucolor,
    restrict y to domain=1:inf,
    ]
\addplot table[x=dim,y=222,col sep=comma] {plotdata/square_max_10core.csv};
\addplot table[x=dim,y=232,col sep=comma] {plotdata/square_max_10core.csv};
\addplot table[x=dim,y=234,col sep=comma] {plotdata/square_max_10core.csv};
\addplot table[x=dim,y=243,col sep=comma] {plotdata/square_max_10core.csv};
\addplot table[x=dim,y=252,col sep=comma] {plotdata/square_max_10core.csv};
\addplot table[x=dim,y=322,col sep=comma] {plotdata/square_max_10core.csv};
\addplot table[x=dim,y=323,col sep=comma] {plotdata/square_max_10core.csv};
\addplot table[x=dim,y=324,col sep=comma] {plotdata/square_max_10core.csv};
\addplot table[x=dim,y=332,col sep=comma] {plotdata/square_max_10core.csv};
\addplot table[x=dim,y=333,col sep=comma] {plotdata/square_max_10core.csv};
\addplot table[x=dim,y=336,col sep=comma] {plotdata/square_max_10core.csv};
\addplot table[x=dim,y=342,col sep=comma] {plotdata/square_max_10core.csv};
\addplot table[x=dim,y=343,col sep=comma] {plotdata/square_max_10core.csv};
\addplot table[x=dim,y=353,col sep=comma] {plotdata/square_max_10core.csv};
\addplot table[x=dim,y=363,col sep=comma] {plotdata/square_max_10core.csv};
\addplot table[x=dim,y=422,col sep=comma] {plotdata/square_max_10core.csv};
\addplot table[x=dim,y=423,col sep=comma] {plotdata/square_max_10core.csv};
\addplot table[x=dim,y=424,col sep=comma] {plotdata/square_max_10core.csv};
\addplot table[x=dim,y=432,col sep=comma] {plotdata/square_max_10core.csv};
\addplot table[x=dim,y=433,col sep=comma] {plotdata/square_max_10core.csv};
\addplot table[x=dim,y=442,col sep=comma] {plotdata/square_max_10core.csv};
\addplot table[x=dim,y=522,col sep=comma] {plotdata/square_max_10core.csv};
\addplot table[x=dim,y=633,col sep=comma] {plotdata/square_max_10core.csv};
\addplot table[x=dim,y=blis,col sep=comma]{plotdata/square_max_10core.csv};
\addplot table[x=dim,y=mkl,col sep=comma] {plotdata/square_max_10core.csv};
\end{axis}
\end{tikzpicture}
\begin{tikzpicture}[scale=.9]
\begin{axis}[
    title={$m = k = n$, 10 core, \emph{Reference}},
    xlabel={ $m = k = n$ },
    ylabel={\emph{Effective} GFLOPS ($ 2 \cdot  m \cdot  n \cdot k / time $)},
    xmin=0,
    xmax=12000,
    ymin=150,
    ymax=255,
    xtick={1000,2000,3000,4000,5000,6000,7000,8000,9000,10000,11000,12000},
    ytick={50, 100, 150, 200, 248},
    scaled x ticks=false,
    scaled x ticks=base 10:-3,
    grid=major,
    axis background/.style={fill=lightgray!20},
    mark size=0.8pt,
    cycle list name=jianyucolor,
    restrict y to domain=1:inf,
    ]
\addplot table[x=dim,y=222,col sep=comma] {plotdata/austin_square_max_10core.csv};
\addplot table[x=dim,y=232,col sep=comma] {plotdata/austin_square_max_10core.csv};
\addplot table[x=dim,y=234,col sep=comma] {plotdata/austin_square_max_10core.csv};
\addplot table[x=dim,y=243,col sep=comma] {plotdata/austin_square_max_10core.csv};
\addplot table[x=dim,y=252,col sep=comma] {plotdata/austin_square_max_10core.csv};
\addplot table[x=dim,y=322,col sep=comma] {plotdata/austin_square_max_10core.csv};
\addplot table[x=dim,y=323,col sep=comma] {plotdata/austin_square_max_10core.csv};
\addplot table[x=dim,y=324,col sep=comma] {plotdata/austin_square_max_10core.csv};
\addplot table[x=dim,y=332,col sep=comma] {plotdata/austin_square_max_10core.csv};
\addplot table[x=dim,y=333,col sep=comma] {plotdata/austin_square_max_10core.csv};
\addplot table[x=dim,y=336,col sep=comma] {plotdata/austin_square_max_10core.csv};
\addplot table[x=dim,y=342,col sep=comma] {plotdata/austin_square_max_10core.csv};
\addplot table[x=dim,y=343,col sep=comma] {plotdata/austin_square_max_10core.csv};
\addplot table[x=dim,y=353,col sep=comma] {plotdata/austin_square_max_10core.csv};
\addplot table[x=dim,y=363,col sep=comma] {plotdata/austin_square_max_10core.csv};
\addplot table[x=dim,y=422,col sep=comma] {plotdata/austin_square_max_10core.csv};
\addplot table[x=dim,y=423,col sep=comma] {plotdata/austin_square_max_10core.csv};
\addplot table[x=dim,y=424,col sep=comma] {plotdata/austin_square_max_10core.csv};
\addplot table[x=dim,y=432,col sep=comma] {plotdata/austin_square_max_10core.csv};
\addplot table[x=dim,y=433,col sep=comma] {plotdata/austin_square_max_10core.csv};
\addplot table[x=dim,y=442,col sep=comma] {plotdata/austin_square_max_10core.csv};
\addplot table[x=dim,y=522,col sep=comma] {plotdata/austin_square_max_10core.csv};
\addplot table[x=dim,y=633,col sep=comma] {plotdata/austin_square_max_10core.csv};
\addplot table[x=dim,y=blis,col sep=comma]{plotdata/austin_square_max_10core.csv};
\addplot table[x=dim,y=mkl,col sep=comma] {plotdata/austin_square_max_10core.csv};
\end{axis}
\end{tikzpicture}
\begin{tikzpicture}[scale=.9]
\begin{axis}[
    title={$m = n = 14400$, 10 core, \emph{Ours}},
    xlabel={ $k$ },
    ylabel={\emph{Effective} GFLOPS ($ 2 \cdot  m \cdot  n \cdot k / time $)},
    xmin=0,
    xmax=12000,
    ymin=150,
    ymax=255,
    xtick={1000,2000,3000,4000,5000,6000,7000,8000,9000,10000,11000,12000},
    ytick={50, 100, 150, 200, 248},
    scaled x ticks=false,
    scaled x ticks=base 10:-3,
    grid=major,
    axis background/.style={fill=lightgray!20},
    mark size=0.8pt,
    cycle list name=jianyucolor,
    restrict y to domain=1:inf,
    ]
\addplot table[x=dim,y=222,col sep=comma] {plotdata/rankk_max_10core.csv};
\addplot table[x=dim,y=232,col sep=comma] {plotdata/rankk_max_10core.csv};
\addplot table[x=dim,y=234,col sep=comma] {plotdata/rankk_max_10core.csv};
\addplot table[x=dim,y=243,col sep=comma] {plotdata/rankk_max_10core.csv};
\addplot table[x=dim,y=252,col sep=comma] {plotdata/rankk_max_10core.csv};
\addplot table[x=dim,y=322,col sep=comma] {plotdata/rankk_max_10core.csv};
\addplot table[x=dim,y=323,col sep=comma] {plotdata/rankk_max_10core.csv};
\addplot table[x=dim,y=324,col sep=comma] {plotdata/rankk_max_10core.csv};
\addplot table[x=dim,y=332,col sep=comma] {plotdata/rankk_max_10core.csv};
\addplot table[x=dim,y=333,col sep=comma] {plotdata/rankk_max_10core.csv};
\addplot table[x=dim,y=336,col sep=comma] {plotdata/rankk_max_10core.csv};
\addplot table[x=dim,y=342,col sep=comma] {plotdata/rankk_max_10core.csv};
\addplot table[x=dim,y=343,col sep=comma] {plotdata/rankk_max_10core.csv};
\addplot table[x=dim,y=353,col sep=comma] {plotdata/rankk_max_10core.csv};
\addplot table[x=dim,y=363,col sep=comma] {plotdata/rankk_max_10core.csv};
\addplot table[x=dim,y=422,col sep=comma] {plotdata/rankk_max_10core.csv};
\addplot table[x=dim,y=423,col sep=comma] {plotdata/rankk_max_10core.csv};
\addplot table[x=dim,y=424,col sep=comma] {plotdata/rankk_max_10core.csv};
\addplot table[x=dim,y=432,col sep=comma] {plotdata/rankk_max_10core.csv};
\addplot table[x=dim,y=433,col sep=comma] {plotdata/rankk_max_10core.csv};
\addplot table[x=dim,y=442,col sep=comma] {plotdata/rankk_max_10core.csv};
\addplot table[x=dim,y=522,col sep=comma] {plotdata/rankk_max_10core.csv};
\addplot table[x=dim,y=633,col sep=comma] {plotdata/rankk_max_10core.csv};
\addplot table[x=dim,y=blis,col sep=comma]{plotdata/rankk_max_10core.csv};
\addplot table[x=dim,y=mkl,col sep=comma] {plotdata/rankk_max_10core.csv};
\end{axis}
\end{tikzpicture}
\begin{tikzpicture}[scale=.9]
\begin{axis}[
    title={$m = n = 14400$, 10 core, \emph{Reference}},
    xlabel={ $k$ },
    ylabel={\emph{Effective} GFLOPS ($ 2 \cdot  m \cdot  n \cdot k / time $)},
    xmin=0,
    xmax=12000,
    ymin=150,
    ymax=255,
    xtick={1000,2000,3000,4000,5000,6000,7000,8000,9000,10000,11000,12000},
    ytick={50, 100, 150, 200, 248},
    scaled x ticks=false,
    scaled x ticks=base 10:-3,
    grid=major,
    axis background/.style={fill=lightgray!20},
    mark size=0.8pt,
    cycle list name=jianyucolor,
    restrict y to domain=1:inf,
    legend style={
        at={(0.99,0.01)},
        anchor=south east,
        legend columns=4,
        font=\tiny,
        rounded corners=1pt,
        nodes={scale=0.8, transform shape},
    },
    legend entries = {
        $ \langle 2, 2, 2 \rangle $\\
        $ \langle 2, 3, 2 \rangle $\\ 
        $ \langle 2, 3, 4 \rangle $\\ 
        $ \langle 2, 4, 3 \rangle $\\ 
        $ \langle 2, 5, 2 \rangle $\\ 
        $ \langle 3, 2, 2 \rangle $\\ 
        $ \langle 3, 2, 3 \rangle $\\ 
        $ \langle 3, 2, 4 \rangle $\\ 
        $ \langle 3, 3, 2 \rangle $\\ 
        $ \langle 3, 3, 3 \rangle $\\ 
        $ \langle 3, 3, 6 \rangle $\\ 
        $ \langle 3, 4, 2 \rangle $\\ 
        $ \langle 3, 4, 3 \rangle $\\ 
        $ \langle 3, 5, 3 \rangle $\\ 
        $ \langle 3, 6, 3 \rangle $\\ 
        $ \langle 4, 2, 2 \rangle $\\ 
        $ \langle 4, 2, 3 \rangle $\\ 
        $ \langle 4, 2, 4 \rangle $\\ 
        $ \langle 4, 3, 2 \rangle $\\ 
        $ \langle 4, 3, 3 \rangle $\\ 
        $ \langle 4, 4, 2 \rangle $\\ 
        $ \langle 5, 2, 2 \rangle $\\ 
        $ \langle 6, 3, 3 \rangle $\\
        $ BLIS $\\
        $ MKL $\\
    },
    ]
\addplot table[x=dim,y=222,col sep=comma] {plotdata/austin_rankk_max_10core.csv};
\addplot table[x=dim,y=232,col sep=comma] {plotdata/austin_rankk_max_10core.csv};
\addplot table[x=dim,y=234,col sep=comma] {plotdata/austin_rankk_max_10core.csv};
\addplot table[x=dim,y=243,col sep=comma] {plotdata/austin_rankk_max_10core.csv};
\addplot table[x=dim,y=252,col sep=comma] {plotdata/austin_rankk_max_10core.csv};
\addplot table[x=dim,y=322,col sep=comma] {plotdata/austin_rankk_max_10core.csv};
\addplot table[x=dim,y=323,col sep=comma] {plotdata/austin_rankk_max_10core.csv};
\addplot table[x=dim,y=324,col sep=comma] {plotdata/austin_rankk_max_10core.csv};
\addplot table[x=dim,y=332,col sep=comma] {plotdata/austin_rankk_max_10core.csv};
\addplot table[x=dim,y=333,col sep=comma] {plotdata/austin_rankk_max_10core.csv};
\addplot table[x=dim,y=336,col sep=comma] {plotdata/austin_rankk_max_10core.csv};
\addplot table[x=dim,y=342,col sep=comma] {plotdata/austin_rankk_max_10core.csv};
\addplot table[x=dim,y=343,col sep=comma] {plotdata/austin_rankk_max_10core.csv};
\addplot table[x=dim,y=353,col sep=comma] {plotdata/austin_rankk_max_10core.csv};
\addplot table[x=dim,y=363,col sep=comma] {plotdata/austin_rankk_max_10core.csv};
\addplot table[x=dim,y=422,col sep=comma] {plotdata/austin_rankk_max_10core.csv};
\addplot table[x=dim,y=423,col sep=comma] {plotdata/austin_rankk_max_10core.csv};
\addplot table[x=dim,y=424,col sep=comma] {plotdata/austin_rankk_max_10core.csv};
\addplot table[x=dim,y=432,col sep=comma] {plotdata/austin_rankk_max_10core.csv};
\addplot table[x=dim,y=433,col sep=comma] {plotdata/austin_rankk_max_10core.csv};
\addplot table[x=dim,y=442,col sep=comma] {plotdata/austin_rankk_max_10core.csv};
\addplot table[x=dim,y=522,col sep=comma] {plotdata/austin_rankk_max_10core.csv};
\addplot table[x=dim,y=633,col sep=comma] {plotdata/austin_rankk_max_10core.csv};
\addplot table[x=dim,y=blis,col sep=comma]{plotdata/austin_rankk_max_10core.csv};
\addplot table[x=dim,y=mkl,col sep=comma] {plotdata/austin_rankk_max_10core.csv};
\end{axis}
\end{tikzpicture}
\begin{tikzpicture}[scale=.9]
\begin{axis}[
    title={$k = 1024$, 10 core, \emph{Ours}},
    xlabel={ $m = n$ },
    ylabel={\emph{Effective} GFLOPS ($ 2 \cdot  m \cdot  n \cdot k / time $)},
    xmin=0,
    xmax=12000,
    ymin=150,
    ymax=255,
    xtick={1000,2000,3000,4000,5000,6000,7000,8000,9000,10000,11000,12000},
    ytick={50, 100, 150, 200, 248},
    scaled x ticks=false,
    scaled x ticks=base 10:-3,
    grid=major,
    axis background/.style={fill=lightgray!20},
    mark size=0.8pt,
    cycle list name=jianyucolor,
    restrict y to domain=1:inf,
    ]
\addplot table[x=dim,y=222,col sep=comma] {plotdata/fixk_max_10core.csv};
\addplot table[x=dim,y=232,col sep=comma] {plotdata/fixk_max_10core.csv};
\addplot table[x=dim,y=234,col sep=comma] {plotdata/fixk_max_10core.csv};
\addplot table[x=dim,y=243,col sep=comma] {plotdata/fixk_max_10core.csv};
\addplot table[x=dim,y=252,col sep=comma] {plotdata/fixk_max_10core.csv};
\addplot table[x=dim,y=322,col sep=comma] {plotdata/fixk_max_10core.csv};
\addplot table[x=dim,y=323,col sep=comma] {plotdata/fixk_max_10core.csv};
\addplot table[x=dim,y=324,col sep=comma] {plotdata/fixk_max_10core.csv};
\addplot table[x=dim,y=332,col sep=comma] {plotdata/fixk_max_10core.csv};
\addplot table[x=dim,y=333,col sep=comma] {plotdata/fixk_max_10core.csv};
\addplot table[x=dim,y=336,col sep=comma] {plotdata/fixk_max_10core.csv};
\addplot table[x=dim,y=342,col sep=comma] {plotdata/fixk_max_10core.csv};
\addplot table[x=dim,y=343,col sep=comma] {plotdata/fixk_max_10core.csv};
\addplot table[x=dim,y=353,col sep=comma] {plotdata/fixk_max_10core.csv};
\addplot table[x=dim,y=363,col sep=comma] {plotdata/fixk_max_10core.csv};
\addplot table[x=dim,y=422,col sep=comma] {plotdata/fixk_max_10core.csv};
\addplot table[x=dim,y=423,col sep=comma] {plotdata/fixk_max_10core.csv};
\addplot table[x=dim,y=424,col sep=comma] {plotdata/fixk_max_10core.csv};
\addplot table[x=dim,y=432,col sep=comma] {plotdata/fixk_max_10core.csv};
\addplot table[x=dim,y=433,col sep=comma] {plotdata/fixk_max_10core.csv};
\addplot table[x=dim,y=442,col sep=comma] {plotdata/fixk_max_10core.csv};
\addplot table[x=dim,y=522,col sep=comma] {plotdata/fixk_max_10core.csv};
\addplot table[x=dim,y=633,col sep=comma] {plotdata/fixk_max_10core.csv};
\addplot table[x=dim,y=blis,col sep=comma]{plotdata/fixk_max_10core.csv};
\addplot table[x=dim,y=mkl,col sep=comma] {plotdata/fixk_max_10core.csv};
\end{axis}
\end{tikzpicture}
\begin{tikzpicture}[scale=.9]
\begin{axis}[
    title={$k = 1024$, 10 core, \emph{Reference}},
    xlabel={ $m = n$ },
    ylabel={\emph{Effective} GFLOPS ($ 2 \cdot  m \cdot  n \cdot k / time $)},
    xmin=0,
    xmax=12000,
    ymin=150,
    ymax=255,
    xtick={1000,2000,3000,4000,5000,6000,7000,8000,9000,10000,11000,12000},
    ytick={50, 100, 150, 200, 248},
    scaled x ticks=false,
    scaled x ticks=base 10:-3,
    grid=major,
    axis background/.style={fill=lightgray!20},
    mark size=0.8pt,
    cycle list name=jianyucolor,
    restrict y to domain=1:inf,
    ]
\addplot table[x=dim,y=222,col sep=comma] {plotdata/austin_fixk_max_10core.csv};
\addplot table[x=dim,y=232,col sep=comma] {plotdata/austin_fixk_max_10core.csv};
\addplot table[x=dim,y=234,col sep=comma] {plotdata/austin_fixk_max_10core.csv};
\addplot table[x=dim,y=243,col sep=comma] {plotdata/austin_fixk_max_10core.csv};
\addplot table[x=dim,y=252,col sep=comma] {plotdata/austin_fixk_max_10core.csv};
\addplot table[x=dim,y=322,col sep=comma] {plotdata/austin_fixk_max_10core.csv};
\addplot table[x=dim,y=323,col sep=comma] {plotdata/austin_fixk_max_10core.csv};
\addplot table[x=dim,y=324,col sep=comma] {plotdata/austin_fixk_max_10core.csv};
\addplot table[x=dim,y=332,col sep=comma] {plotdata/austin_fixk_max_10core.csv};
\addplot table[x=dim,y=333,col sep=comma] {plotdata/austin_fixk_max_10core.csv};
\addplot table[x=dim,y=336,col sep=comma] {plotdata/austin_fixk_max_10core.csv};
\addplot table[x=dim,y=342,col sep=comma] {plotdata/austin_fixk_max_10core.csv};
\addplot table[x=dim,y=343,col sep=comma] {plotdata/austin_fixk_max_10core.csv};
\addplot table[x=dim,y=353,col sep=comma] {plotdata/austin_fixk_max_10core.csv};
\addplot table[x=dim,y=363,col sep=comma] {plotdata/austin_fixk_max_10core.csv};
\addplot table[x=dim,y=422,col sep=comma] {plotdata/austin_fixk_max_10core.csv};
\addplot table[x=dim,y=423,col sep=comma] {plotdata/austin_fixk_max_10core.csv};
\addplot table[x=dim,y=424,col sep=comma] {plotdata/austin_fixk_max_10core.csv};
\addplot table[x=dim,y=432,col sep=comma] {plotdata/austin_fixk_max_10core.csv};
\addplot table[x=dim,y=433,col sep=comma] {plotdata/austin_fixk_max_10core.csv};
\addplot table[x=dim,y=442,col sep=comma] {plotdata/austin_fixk_max_10core.csv};
\addplot table[x=dim,y=522,col sep=comma] {plotdata/austin_fixk_max_10core.csv};
\addplot table[x=dim,y=633,col sep=comma] {plotdata/austin_fixk_max_10core.csv};
\addplot table[x=dim,y=blis,col sep=comma]{plotdata/austin_fixk_max_10core.csv};
\addplot table[x=dim,y=mkl,col sep=comma] {plotdata/austin_fixk_max_10core.csv};
\end{axis}
\end{tikzpicture}
\caption{Performance of the best implementation of our generated FMM code and reference implementations \cite{Benson15} on one socket (10 core).
Top row: our implementations; Bottom row: reference implementations from \cite{Benson15} (linked with Intel MKL).
Left column: \SQUARE{}; Middle column: \RANKK{}; Right column: \FIXK{}. 
}
\label{fig:10core}
\end{figure*}

\subsection{Sequential and parallel performance}

Results when using a single core are presented in Figures \ref{tab:fmm_alg}, \ref{fig:rankk_1level_1core}, and \ref{fig:2level_abc_1core}.
Our generated \ABCFMM{} implementation outperforms \ABXFMM{}/\XXXFMM{} and reference implementations from \cite{Benson15}
for rank-k updates (when $ k $ is small).
For very large square matrices, our generated \ABXFMM{} or \XXXFMM{} can achieve competitive performance
with reference implementations \cite{Benson15} that is linked with Intel MKL.
These experiments support the validity of our model.




\figref{fig:10core} reports performance results for ten cores within the same socket.
Memory bandwidth contention impacts the performance of various FMM when using many cores.
Nonetheless we still observe the speedup of FMM over \gemm{}.
For smaller matrices and special shapes such as rank-k updates, 
our generated implementations achieve better performance than reference implementations \cite{Benson15}.

%
%
%
%


\NoShow{
\subsection{Many-core performance}

\begin{figure}[htp!]
\center 
\begin{tikzpicture}[scale=1.0]
\begin{axis}[
    title={$m = n = 14400$, ABC, 64 core},
    xlabel={ $ k $ },
    ylabel={\emph{Effective} GFLOPS ($ 2 \cdot  m \cdot  n \cdot k / time $)},
    xmin=0,
    xmax=7000,
    ymin=600,
    ymax=2000,
    xtick={1000,2000,3000,4000,5000,6000,7000,8000,9000,10000,11000,12000,13000,14000,15000},
    ytick={0, 200, 400, 600, 800, 1000, 1200, 1400, 1600, 1800, 2000},
    scaled x ticks=false,
    scaled x ticks=base 10:-3,
    grid=major,
    axis background/.style={fill=lightgray!20},
    mark size=0.8pt,
    cycle list name=jianyu3color,
    legend style={
        at={(0.99,0.01)},
        anchor=south east,
        legend columns=1,
        font=\tiny,
        rounded corners=1pt,
        nodes={scale=0.8, transform shape},
    },
    legend entries = {
        $ TBLIS    $\\
        $ MKL     $\\
        $ \langle 2, 2, 2 \rangle, 1~level$\\
        $ \langle 2, 2, 2 \rangle, 2~level$\\
        $ \langle 3, 2, 3 \rangle, 1~level$\\
    },
    ]
\addplot table[x=dim,y=blis,col sep=comma]  {plotdata/knl_perf_result.csv};
\addplot table[x=dim,y=mkl,col sep=comma]   {plotdata/knl_perf_result.csv};
\addplot table[x=dim,y=222_1,col sep=comma] {plotdata/knl_perf_result.csv};
\addplot table[x=dim,y=222_2,col sep=comma] {plotdata/knl_perf_result.csv};
\addplot table[x=dim,y=323_1,col sep=comma] {plotdata/knl_perf_result.csv};
\end{axis}
\end{tikzpicture}
\caption{KNL performance result.}
\label{fig:knl}
\end{figure}


\subsubsection*{Implementation}

The generated code is in {\tt C} and  {\tt AVX512} assembly, compiled with the Intel\rr{} {\tt C} compiler version 17.0.0 with optimization flag {\tt -O3 -xMIC-AVX512}.

Our generated micro-kernel code is based on TBLIS project\footnote{https://github.com/devinamatthews/blis/tree/knl}.
The generated code parallelize the $1^{st}$ and $3^{rd}$ loop around the micro-kernel with $ 4 \times 16 $ parallelization scheme.

\subsubsection*{Target architecture}

We run the KNL performance experiments on a 
Intel Xeon Phi CPU 7210 coprocessor (1.3 Ghz).
This coprocessor
has a peak performance of 2.66 TFLOPS (for 64 cores) 
\footnote{ 
$ 2660 = 64 \times 1.3 \times 32 $, assuming 2 512-bit VPUs dual issuing {\tt FMA}.
However, MKL reaches 1.8 TFLOPS, 69\% efficiency.
KNL may have a different AVX base frequency, and the clockrate may
drop to 1.0 GHz.
}
16 GB of MCDRAM with a peak bandwidth of 450 GB/s. It has 1 MB L2 cache, shared by every two cores in the same tile, but no L3 cache.


We choose the parameters $n_R=8$, $m_R=24$, $k_C=336$, $n_C=14400$ and $m_C=120$. This makes the size of the packing buffer
$ \widetilde A_i $ 323 KB and 
$ \widetilde B_p $ 38.7 MB, which fits L2 cache and MCDRAM/main memory separately (no L3 cache on KNL).  These choices are consistent with those used by TBLIS.
KNL processor is configured with flat-quadrant mode, and we use {\tt memkind} to allocate memory on MCDRAM as high bandth memory. To ensure the load balance and thread affinity on all 64 cores, we use
{\tt KMP\_AFFINITY=scatter} setting.

\subsubsection*{Results}
\figref{fig:knl} reports the performance of one-level/two-level $ \langle 2,2,2 \rangle $
and one-level $ \langle 3,2,3 \rangle $.

}

\section{Conclusion} \label{s:conclusion}

We \FromTo{developed}{have discussed} a code generator framework \FromTo{which}{that} can automatically 
implement families of FMM algorithms for \FromTo{each}{}Strassen-like fast matrix multiplication algorithms.
This code generator expresses the composition of multi-level FMM algorithms as Kronecker products.
It incorporates the matrix summations that must be performed for FMM into the
inherent packing and micro-kernel operations inside \gemm{}, avoiding extra workspace requirement and
reducing the overhead of memory movement.
Importantly, it \FromTo{can embed}{generates} an
accurate performance model to guide the selection of a FMM implementation as a function of problem size and shape,
facilitating the creation of poly-algorithms that select the best algorithm for a problem size.
Comparing with state-of-the-art results,
we observe a significant performance improvement for smaller matrices and special matrix multiplication shapes
such as rank-k updates, without the need for exhaustive empirical search.

There are a number of avenues for future work:
\begin{itemize}
\item Task parallelism and various parallel schemes
are proposed in the recent literature
\cite{D'alberto:2011:EPM:2049662.2049664,Benson15}.
We need to pursue how our techniques compare to these
and how to combine these with our advances. It may be possible to
utilize our performance model to help with task scheduling.

\item Finding the new FMM algorithms by searching the coefficient matrix
$ \llbracket {U}, {V}, {W} \rrbracket $ is an NP-hard problem \cite{Knuth1997}.
It may be possible to prune branches
with the performance model as the cost function during the search process.


\item 
In \cite{Strassen:SC16}, it is shown that Intel Xeon Phi coprocessor (KNC) can benefit from ABC variation of $ \langle 2,2,2 \rangle $ \strassen{}.
It may be possible to get performance benefit by porting our code generator to generate variations of FMM implementations for many-core architecture such as
second-generation Intel Xeon Phi coprocessor (KNL).

\end{itemize}



\section*{Additional information}

Additional information regarding BLIS and related projects can be found at 
\begin{center}
{\tt http://shpc.ices.utexas.edu}
\end{center}

\section*{Acknowledgments}

This work was sponsored in part by the National Science Foundation under grant number ACI-1550493, by Intel Corporation through an Intel Parallel Computing Center grant, and by a gift from Qualcomm. Access to the Maverick supercomputers administered by TACC is gratefully acknowledged.
Jianyu Huang was supported by a summer fellowship from Graduate School of UT Austin.
DAM is an Arnold O. Beckman Postdoctoral Fellow.
We thank the rest of the SHPC team ({\tt http://shpc.ices.utexas.edu}) for their supports.

{\em Any opinions, findings, and conclusions or recommendations expressed in this material are those of the author(s) and do not necessarily reflect the views of the National Science Foundation. }

%


\bibliographystyle{IEEEtran}

\bibliography{biblio}


\end{document}